\documentclass[draftclsnofoot,onecolumn,12pt]{IEEEtran}
\IEEEoverridecommandlockouts
\UseRawInputEncoding
\usepackage{subfigure} 
\usepackage{graphicx}

\usepackage{amsmath,graphicx,amssymb,mathtools,bm}
\usepackage{subfigure}
\usepackage{hyperref}
\usepackage{cite}
\usepackage{amsmath,amssymb,amsfonts}
\usepackage{amsthm}
\usepackage{algorithmic}
\usepackage{graphicx}
\usepackage{textcomp}
\usepackage{xcolor}
\usepackage{verbatim}  
\usepackage{graphicx}  
\usepackage{bm}  
\usepackage{mathrsfs} 
\usepackage{algorithm} 
\usepackage{algorithmic} 
\usepackage{booktabs}
\usepackage{textcomp}  
\usepackage{multirow}  
\usepackage{lettrine}   
\usepackage{graphicx}  
\usepackage{subeqnarray}

\usepackage{cases}
\usepackage{color}  
\usepackage{amsmath}
\usepackage{amssymb}
\usepackage{stfloats}
\usepackage{caption}
\usepackage{subfigure}

\usepackage{titlesec}
\DeclareMathSizes{12}{10.3}{7}{5}
\usepackage[linesnumbered,ruled,vlined,algo2e]{algorithm2e}

\def\BibTeX{{\rm B\kern-.05em{\sc i\kern-.025em b}\kern-.08em
    T\kern-.1667em\lower.7ex\hbox{E}\kern-.125emX}}

\makeatletter
\renewcommand{\maketag@@@}[1]{\hbox{\m@th\normalsize\normalfont#1}}%
\makeatother

 \newtheorem{Proposition}{\bf Proposition}
 \newtheorem{remark}{Remark}
 \newenvironment{Proof}{{\indent \it Proof:}}{\hfill $\blacksquare$\par}

\begin{document}
	\newcommand{\tabincell}[2]{\begin{tabular}{@{}#1@{}}#2\end{tabular}} 

\title{Near-Field Channel Estimation for Extremely Large-Scale Reconfigurable Intelligent Surface (XL-RIS)-Aided Wideband mmWave Systems 
}

\author{Songjie Yang, Chenfei Xie, Wanting Lyu, Boyu Ning, \IEEEmembership{Member,~IEEE}\\ Zhongpei Zhang, \IEEEmembership{Member,~IEEE}, and Chau Yuen, \IEEEmembership{Fellow,~IEEE}

\thanks{Songjie Yang, Chenfei Xie, Wanting Lyu, Boyu Ning,
	 and Zhongpei Zhang are with the National Key Laboratory of Science and Technology on Communications, University of Electronic Science and Technology of China, Chengdu 611731, China. 
	(e-mail:
	yangsongjie@std.uestc.edu.cn;201911220505@std.uestc.edu.cn;
	lyuwanting@yeah.net;boydning@outlook.com;
	zhangzp@uestc.edu.cn).
	
	Chau Yuen is with the School of Electrical and Electronics Engineering, Nanyang Technological University (e-mail: chau.yuen@ntu.edu.sg).}}


\maketitle

\begin{abstract} 
Near-field communications present new opportunities over near-field channels,
however, the spherical wavefront propagation makes near-field signal processing challenging. In this context, this paper proposes efficient near-field channel estimation methods for wideband MIMO mmWave systems with the aid of extremely large-scale reconfigurable intelligent surfaces (XL-RIS). For the wideband signals reflected by the analog RIS, we characterize their near-field beam squint effect in both angle and distance domains.
 Based on the mathematical analysis of the near-field beam patterns over all frequencies, a wideband spherical-domain dictionary is constructed by minimizing the coherence of two arbitrary beams. In light of this, we formulate a two-dimensional compressive sensing problem to recover the channel parameter based on the spherical-domain sparsity of mmWave channels. To this end,  we present a correlation coefficient-based atom matching method within our proposed multi-frequency parallelizable subspace recovery framework for efficient solutions.
  Additionally, we propose a two-dimensional oracle estimator as a benchmark and derive its lower bound across all subcarriers. Our findings emphasize the significance of system hyperparameters and the sensing matrix of each subcarrier in determining the accuracy of the estimation.  Finally, numerical results show that our proposed method achieves considerable performance compared with the lower bound and has a time complexity linear to the number of RIS elements. 

\end{abstract}
\vspace{-0.55em}
\section{Introduction}

\lettrine[lines=2]{W} 
ireless communication systems have evolved rapidly over the past few decades, with each new generation of technology bringing a significant increase in capacity and data rates. However, as the demand for wireless communications and data access continues to grow, it has become clear that current wireless technologies will eventually reach their limit. To overcome this challenge, researchers have turned to  promising new technologies such as extremely large-scale multiple-input multiple-output (XL-MIMO) \cite{XL1}, holographic MIMO \cite{HMIMO}, and orbital angular momentum MIMO \cite{OAM}. These technologies have the potential to improve the reliability and quality of wireless services, particularly in high-density urban areas, where traditional wireless technologies struggle to keep with demand.

The potential of XL-MIMO technology to significantly increase spectral efficiency and enhance localization has been demonstrated by the authors of \cite{XL1,XL2,XL3,XL4}, making it an enticing option for next-generation wireless communication and sensing systems. However, the use of XL-arrays poses a challenge due to the invalidation of the planar wavefront assumption in the near-field region \cite{near-field1}. In its place, the  spherical wavefront assumption should be considered for near-field channel characterization, which depends not only on the spatial angle but also on the distance between the array and the source or scatter. As a result, channel estimation and modeling in the near-field region are even more demanding than in the far-field region, requiring greater precision and accuracy.

On the other hand, reconfigurable intelligent surface (RIS), capable of reflecting, refracting and manipulating incoming electromagnetic waves \cite{SMART}, is particularly promising as it allows to overcome some of the propagation challenges associated with the mmWave/THz spectrum \cite{RIS1,RIS2}. 
There are two distinct types of RIS that are used in wireless communication systems: near-field and far-field RIS.
Far-field RIS operates in the far-field region, which is defined as the region, where the radiated field is essentially plane-wave-like and the distance from the RIS is much larger than the size of the RIS. In this case, the RIS can be modeled as modifying the radiation pattern of the radiating source, and the primary goal is to steer the beam in a desired direction or shape. 
Over these years, various fundamental topics have been studied for far-field RIS \cite{FRIS1,FRIS2,FRIS3,FRIS4,FRIS6,RIS3,NB1,NB2},
including 
channel estimation, localization, and passive/active beamforming.  

By leveraging the advantages of XL-MIMO technology, several studies have started exploring near-field XL-RIS techniques that entail deploying numerous small, low-cost, and passive/active elements in close proximity to either the transmitter or receiver. In this region, the RIS can be viewed as a thin surface that modifies the local electromagnetic environment. The primary goal of near-field RIS is to optimize the wireless signal quality at a specific location.
However, the near electromagnetic field exhibits complex behavior that varies significantly depending on the distance between the transmitter/receiver and the RIS.
Despite the challenges involved in near-field region, it is an important area of research that has the potential to improve the performance of wireless communication systems in the near future.
\vspace{-1em}
\subsection{State-of-the-Art Works}
The research on near-field XL-RIS is still in the preliminary stage. The authors of \cite{XLRIS1,XLRIS2} provided complementary perspectives on the use of large intelligent surfaces and near-field beamforming for improving wireless communications in the near-field region. They both highlight the potential of these techniques to enhance the performance and efficiency of wireless communication systems, and provide insights into the design and optimization of these systems for different communication modes and scenarios. Since the focus of near-field RIS is on the spatial beamspace, several related topics have been studied, including beam training \cite{XL-BT1,XL-BT2,XL-BT3}, localization \cite{XL-loc1,XL-loc2,XL-loc3,XL-loc4,XL-loc5}, and channel estimation \cite{XL-loc3,XL-loc4,XLRCE}. In the near-field region, angular beam training using the planar steering vector is no longer effective, making it necessary to redesign the codebook for beam training with the spherical steering vector. In the Cartesian coordinate system, two near-field codebooks have been designed, namely the near-field uniform codebook \cite{XL-BT1} and the near-field hierarchical codebook \cite{XL-BT1,XL-BT2}. In \cite{XL-BT1}, it was shown that the near-field hierarchical codebook incurs significantly less training overhead than the near-field uniform codebook. Alternatively, the authors of \cite{XL-BT3} utilized the polar-domain codebook \cite{near-CE2} to perform near-field beam training in the polar domain.
On the other hand, near-field localization is a popular topic due to the advantageous spherical wavefront for distance sensing. In this context, various studies have investigated near-field localization with the XL-RIS system. Notably, due to the similarity between compressive channel estimation and localization, references \cite{XL-loc3} and \cite{XL-loc4} investigated near-field localization while also studying near-field channel estimation. In \cite{XLRCE}, the authors investigated near-field channel estimation for the hybrid beamforming system with the aid of the XL-RIS.

However, the above references just considered narrowband systems.
In practical systems such as orthogonal frequency division multiplex (OFDM), wideband signals are commonly used, which will cause a well-known spatial effect called beam squint in phased array antennas. Beam squint occurs due to the change in the angle of incidence of the incoming wave with frequency, and it can have a significant impact on the performance of the wideband system. To mitigate the effects of beam squint, researchers have proposed effective methods for wideband channel estimation and hybrid beamforming, as demonstrated in \cite{BSE5,BSE1,BSE2,BSE4}. However, dealing with beam squint can be even more challenging in near-field XL-RIS systems. Using the polar-domain dictionary proposed in \cite{near-CE2}, researchers in \cite{RIS-NF-BSE1} adopted the Kronecker compressive sensing (KCS) framework to estimate the wideband channel by exploiting common sparsity on each subcarrier. Moreover, the near-field beam squint effect was explored in \cite{RIS-NF-BSE2}, which unveiled two important findings in the polar domain: 1) the mathematical correlation between the beam of the central subcarrier and the beam of other subcarriers, and 2) the variation of beam trajectory with respect to frequency. To address beam squint issues in RIS systems at the hardware level, \cite{RIS-NF-BSE3} proposed a delay adjustable metasurface (DAM) architecture, inspired by the true-time delay unit design of phased array antennas. The DAM architecture was subsequently used by \cite{XLRIS3} and \cite{RIS-NF-BSE2} to mitigate the near-field beam squint effect. However, as the number of time delay units increases, the hardware cost and energy consumption also increase.

 Previous research on near-field XL-RIS with beam squint considered uniform linear arrays (ULAs) for channel estimation or localization. However, in the near-field region, the beam squint effect of uniform planar arrays (UPAs) is reflected in the spherical domain, which presents particular challenges for signal processing such as spherical-domain dictionary design \cite{Spherdic}. Last but not least, the XL-RIS system is sensitive to the algorithm complexity due to the numerous elements.
In order to fully exploit the potential of XL-RIS technology for future communication applications, it is essential to improve channel estimation and other signal processing techniques.

\subsection{Contributions and Novelty}
Taking the aforementioned context into account, this paper focuses on the efficient estimation of near-field wideband channels in XL-RIS systems, capitalizing on both the channel's sparsity in the spherical domain and the common sparsity support across all subcarriers. Our approach involves a thorough analysis of the near-field beam squint effect, the development of a wideband spherical-domain dictionary, the proposal of an efficient CS framework, and the derivation of a lower bound. The details of these endeavors are listed below.
\begin{itemize}
\item  
To begin with, we employ the Fresnel approximation \cite{near-field1} to approximate the spherical array response in terms of the elevation/azimuth angle and distance. By separately analyzing the linear and quadratic phase terms, we establish a mathematical relationship between the desired beam at the central frequency and the corresponding beam at other frequencies, captured by the error function ${\rm erf}(\cdot)$. Our exploration of the near-field beam squint effect in the 3D space enables us to produce a comprehensive visualization of the beam trajectory across the frequency spectrum in the spherical domain. Furthermore, drawing on the error function, we formulate a wideband spherical-domain dictionary that minimizes the coherence between any two arbitrary dictionary atoms.

\item We propose the multi-frequency parallelizable subspace recovery (MMPSR) framework for solving the wideband channel estimation problem, using the designed wideband spherical-domain dictionary. This framework converts the 2D-CS problem into multiple sparse vector recovery problems, with multi-frequency joint processing. It is particularly suitable for recovering multi-parameter and large-scale dictionary problems, outperforming commonly used recovery frameworks such as Kronecker CS. To achieve optimal evaluation of the phase relation between the measured signal and the sensing atom, we propose correlation coefficient (CC)-based atom matching, as opposed to inner product (IN)-based atom matching used in some greedy algorithms like orthogonal matching pursuit (OMP).

\item Finally, we propose 
 a 2D-OLS estimator to serve as a reliable benchmark that any estimation method cannot surpass. Furthermore, we derive its lower bound with the evaluation index of the normalized mean squared error (NMSE). Significantly, this lower bound accounts for multi-frequency channels, highlighting the influence of both system hyperparameters (such as the number of antennas, subcarriers, and channel paths) and the sensing matrix of each subcarrier on estimation performance.
\end{itemize}
\vspace{-1em}
\subsection{Organization and Notations}

The rest of this paper is organized as follows: Section \ref{System Model} describes the wideband near-field channels and the uplink training model.
Section \ref{widebanddic} discusses the near-field beam squint effect and designs the wideband spherical-domain dictionary. Section \ref{NCE} proposes the CS framework to process the 2D recovery problem, and gives an efficient solution with CC-based atom matching. Section \ref{LBD} derives the lower bound for the benchmark of any estimation method. Section \ref{simulation} conducts several experiments to demonstrate our proposed methods' effectiveness. Finally, Section \ref{Con} concludes this paper.

{\emph {Notations}}:
${\left(  \cdot  \right)}^{ *}$, ${\left(  \cdot  \right)}^{ T}$, ${\left(  \cdot  \right)}^{ H}$, and $\left(\cdot\right)^{-1}$ denote conjugate, transpose, conjugate transpose, and inverse, respectively.  $\Vert\cdot\Vert_2$ represent $\ell_0$ norm and $\ell_2$ norm, respectively. 
$\Vert\mathbf{A}\Vert_F$ denotes the Frobenius norm of matrix $\mathbf{A}$. ${\rm Tr}\{\mathbf{A}\}$ denotes the trace of matrix $\mathbf{A}$. $\vert\cdot\vert$ denotes the modulus. Furthermore, $\otimes$ is the Kronecker product. $[\mathbf{a}]_{i}$ and $[\mathbf{A}]_{i,j}$ denote the $i$-th element of vector $\mathbf{a}$, the $(i,j)$-th element of matrix $\mathbf{A}$, respectively. $\rm{vec}(\cdot)$ represents the vectorization operation. $\mathbb{E}\{\cdot
\}$ and $\mathbb{V}\{\cdot\}$ denote the expectation and variance operations , respectively.  $\mathbf{I}_M$ denotes the $M$-by-$M$ identity matrix. Moreover, $\rm{diag}(\mathbf{a})$ is a square diagonal matrix with entries of $\mathbf{a}$ on its diagonal. Finally, $\mathcal{CN}(\mathbf{a},\mathbf{A})$ is the complex Gaussian distribution with mean $\mathbf{a}$ and covariance matrix $\mathbf{A}$.

%

\vspace{-0.55em}

 \begin{figure}
	\centering
	\includegraphics[width = 0.88\textwidth]{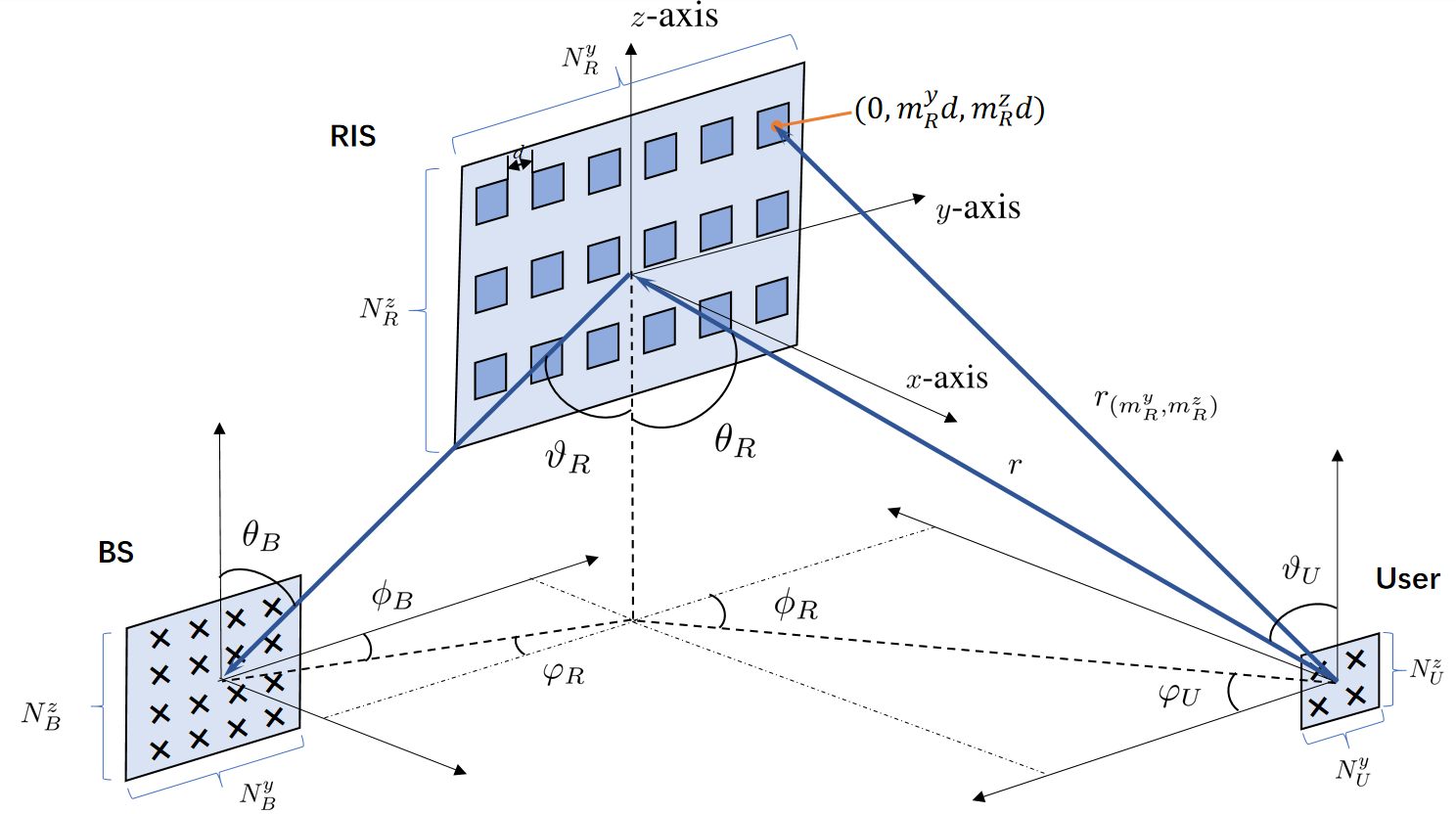}
	\caption{The BS-RIS-user system description. }
	\label{coord}
\end{figure}

\section{Near- and Far-Field Wideband System Model}\label{System Model}
\vspace{-0.15em}

We consider a time division duplexing (TDD) $K$-subcarrier MIMO-OFDM system with the assistance of XL-RIS. Since uplink multiuser channel estimation can be performed in parallel owing to the orthogonal pilot sequence, we just consider an arbitrary user for clarity.
The BS, the RIS, and the user are equipped with a ($N_B=N_B^z\times N_B^y$)-element UPA, a ($N_R=N_R^z\times N_R^y$)-element UPA, and a ($N_U=N_U^z\times N_U^y$)-element UPA, respectively (shown in Fig. \ref{coord}). For simpicity, we assume both the BS and the user are equipped with one radio frequency chain. Moreover, $\mathbf{H}_{\rm B}\in\mathbb{C}^{N_U\times N_R}$, $\mathbf{H}_{\rm U}\in\mathbb{C}^{N_R\times N_B}$, and $\mathbf{H}_{\rm D}\in\mathbb{C}^{N_B\times N_U}$ are the channels between the BS and the RIS, between the RIS and the user, and between the BS and the user\footnote{ This paper ignores the estimation of the direct channel $\mathbf{H}_{\rm D}$ since it can be estimated by turning off the RIS and
 	using conventional estimation methods. }, respectively. The RIS matrix is defined as $\mathbf{V}={\rm diag}(\mathbf{v})\in\mathbb{C}^{N_R\times N_R}$, with $\mathbf{v}\triangleq[e^{jv_1},\cdots,e^{jv_{N_R}}]\in\mathbb{C}^{1\times N_R}$, where $\{v_l\}_{l=1}^{N_R}$ are the phase shifts.  
\vspace{-1em}
 \subsection{Channel Model}
 Considering that the size of the RIS is much larger than that of the user, we utilize a one-side near-field channel model for the RIS-user channel. In the context, the RIS-user channel on the $k$-th subcarrier can be expressed with the Saleh-Valenzuela model:
\begin{equation}\label{H_RU}
	\begin{aligned}
		\mathbf{H}_{\rm U}[k]  = \sqrt{\frac{N_RN_U}{P}}	\sum_{p=1}^{P}\beta_p e^{-j2\pi \tau_p f_k} \mathbf{b}_{\rm R}(f_k,r_{{ R},p})\mathbf{a}_{\rm U}^H(f_k,\widetilde{\vartheta}_{{ U},p},\widetilde{\varphi}_{{ U},p}),
	\end{aligned}
\end{equation} 
where we make a far-field assumption at the user end. $P$ is the number of channel paths,
$\{\tau_p\}_{p=1}^P$ and$\{\beta_p\}_{p=1}^P$ are the path delays and complex path gains, respectively. $r_{R,p}$ denotes the distance from the RIS to the $p$-th object (the scatter or the user).   $\widetilde{\vartheta}_{U,p}\triangleq \cos(\vartheta_{U,p})$ and $\widetilde{\varphi}_{U,p}\triangleq \sin(\vartheta_{U,p})\sin(\varphi_{U,p})$ represent the virtual angles, where  $\vartheta_{U,p}$ and $\varphi_{U,p}$
denote the $p$-th path's elevation and azimuth angles from the user to the RIS, respectively. $f_k\triangleq f_c+\frac{f_s}{K}(k-1-\frac{K-1}{2})$, $k\in\{1,\cdots,K\}$, with $f_c$ being the central frequency and $f_s$ being the bandwidth, denotes the frequency of subcarrier $k$. Moreover, $\mathbf{b}_{\rm R}$ and $\mathbf{a}_{\rm U}$ are the spherical array response and planar array response, respectively.
The spherical array response  on subcarrier $k$ is given by 
\begin{equation}\label{bB}
	\begin{aligned}
		\mathbf{b}_{\rm R}(f_k,r)\triangleq	   \frac{1}{\sqrt{N_{\rm R}}}
		\left[ e^{-j\frac{2\pi f_k}{c} (r_{(-M^y_R,-M^z_R)}-r)},\cdots,e^{-j\frac{2\pi f_k}{c} (r_{(M^y_R,M^z_R)}-r)} \right]^T,
	\end{aligned}
\end{equation}
where $M^{z/y}_R\triangleq \frac{N^{z/y}_R-1}{2}$. As shown in Fig. \ref{coord}, the distance between the $(m^y_R,m^z_R)$-th element with vector coordinate $(0,m^y_Rd,m^z_Rd)$ and the object with   $(r\sin(\theta_{ R})\cos(\phi_{ R}),r\sin(\theta_{ R})\sin(\phi_{ R}),r\cos(\theta_{ R}))$ is denoted by
\begin{equation}
	\begin{aligned}
		r_{(m^y_R,m^z_R)}  &\triangleq\left((r\sin(\theta_{ R})\cos(\phi_{ R}))^2+(r\sin(\theta_{ R})\sin(\phi_{ R})-m^y_Rd)^2 
		+(r\cos(\theta_{ R})-m^z_Rd)^2 \right)^{\frac{1}{2}} \\
		&= \sqrt{r^2-2rm^y_Rd\sin(\theta_{ R})\sin(\phi_{ R})+(m^y_Rd)^2-2rm^z_Rd\cos(\theta_{ R})+(m^z_Rd)^2} \\
		& \overset{(a)}{\approx} r-m^y_Rd \sin(\theta_{ R})\sin(\phi_{ R}) - m^z_Rd\cos(\theta_{ R}) +\frac{(m^z_Rd)^2}{2r}(1-\cos^2(\theta_{ R}))+\frac{(m^y_Rd)^2}{2r}(1-\sin^2(\theta_{ R})\sin^2(\phi_{ R})) \\
		& \ \ \ \ 	 -\frac{m^y_Rm^z_Rd^2\sin(\theta_{ R})\sin(\phi_{ R})\cos(\theta_{ R})}{r} + ... ,
	\end{aligned}
\end{equation}
where $(a)$ holds due to $\sqrt{1+x+y}\approx1+\frac{x+y}{2}-\frac{(x+y)^2}{8}$.

With the far-field assumption, the user's planar array response on subcarrier $k$ can be written as
\begin{equation}\label{aB}
	\begin{aligned}
		\mathbf{a}_{\rm U}(f_k,\widetilde{\vartheta}_U,\widetilde{\varphi}_U)\triangleq
		\frac{1}{\sqrt{N_{U}}} \left[e^{-j \frac{2\pi f_kd}{c} (-M_U^z \widetilde{\vartheta}_U-M_U^y \widetilde{\varphi}_U)},\cdots, e^{-j \frac{2\pi f_kd}{c} (M_U^z \widetilde{\vartheta}_U+M_U^y \widetilde{\varphi}_U)}
		\right]^T,
	\end{aligned}
\end{equation}
where $M^{z/y}_U\triangleq \frac{N^{z/y}_U-1}{2}$.

 Due to the long distance between the BS and the RIS, the far-field assumption can be used for simple approximation. The BS-RIS channel on the $k$-th subcarrier is modeled with a LoS path \cite{XL-loc3}: 
\begin{equation}\label{HB}
	\begin{aligned}
	\mathbf{H}_{\rm B}[k]=  \alpha e^{-j2\pi \tau_0 f_k}\mathbf{a}_{\rm B}(f_k,\widetilde{\theta}_{B},\widetilde{\phi}_{B})\mathbf{a}_{\rm R}^H(f_k,\widetilde{\vartheta}_{R},\widetilde{\varphi}_{R}),
	\end{aligned}
\end{equation} 
where $\alpha$ is the complex path gain and $\tau_0$ is the path delay.    $\widetilde{\theta}_{B}\triangleq \cos(\theta_{B})$ and $\widetilde{\phi}_{B}\triangleq \sin(\theta_{B})\sin(\phi_{B})$, in which $\theta_B$ and $\phi_B$
 are the elevation and azimuth angles from the BS to the RIS, respectively.  $\widetilde{\vartheta}_{R}$ and $\widetilde{\varphi}_{R}$ are defined similarly.
 Moreover, $\mathbf{a}_{\rm B/R}$ is the planar array response, which has a similar form of Eqn. (\ref{aB}).  

\vspace{-1em}
\subsection{Uplink Training}
In the TDD system, uplink channel estimation is performed through continuous transmission of pilots by the user and continuous adjustment of the phase matrix by the RIS. Assuming that $Q$ subframes are utilized for uplink training, where each subframe corresponds to one RIS phase matrix, the RIS fixes the phase matrix in each subframe, while the user transmits $N_X$ pilot sequences. As a result, $QN_X$ symbol durations are required for channel estimation of a single user. The $n$-th training beam on subcarrier $k$ is denoted as $\mathbf{f}_n[k]$ with $n=1,\cdots, N_X$, and the $q$-th RIS phase matrix on all subcarriers is represented as $\mathbf{V}_q\triangleq {\rm diag}(\mathbf{v}_q)$, where $q=1,\cdots, Q$. The training model is described as 
\begin{equation}
y_{q,n}[k]=\mathbf{w}^H[k] \mathbf{H}_{\rm B}[k]\mathbf{V}_q\mathbf{H}_{\rm U}[k]\mathbf{f}_n[k]+\mathbf{w}^H[k]\mathbf{n}_{q,n}[k],
\end{equation}
 where $y_{q,n}[k]$ is the received training signal on subcarrier $k$,  $\mathbf{f}_n[k]\triangleq f_{{\rm BB},n}[k]\mathbf{f}_{{\rm RF},n}\in\mathbb{C}^{N_T\times 1}$ is the hybrid precoder vector on subcarrier $k$ such that $\mathbf{f}^H_n[k]\mathbf{f}_n[k]=\sigma_p^2$ with $\sigma_p^2$ denoting the transmit power. $\mathbf{w}[k]\triangleq w_{\rm BB}[k]\mathbf{w}_{\rm RF}\in\mathbb{C}^{N_R\times 1}$ is the hybrid combiner vector on subcarrier $k$, and $\mathbf{n}_{q,n}[k]\in\mathbb{C}^{N_R \times 1}$ is the noise following $\mathcal{CN}(0,\sigma_n^2\mathbf{I}_{N_R})$. $f_{{\rm BB},n}[k]$ and $w_{\rm BB}[k]$ are the baseband coefficients. Note that $\mathbf{f}_{{\rm RF},n}$ and $\mathbf{w}_{\rm RF}$, the analog vectors produced by the UPA, are fixed at different subcarriers due to analog hardware limitations. With an appropriate location deployment of the BS and the RIS, the BS-RIS channel can be approximated as a LoS channel.
 In this way, we just need to estimate the RIS-user channel. Denoting the effective channel $\widetilde{\mathbf{h}}_{\rm B}[k]\triangleq \mathbf{w}^H[k] \mathbf{H}_{\rm B}[k]$ and the effective noise $\widetilde{n}_{q,n}[k]\triangleq \mathbf{w}^H[k]\mathbf{n}_{q,n}[k]$, where $\mathbf{w}[k]$ can be set towards the angle-of-arrival of the BS. Then, we have
 \begin{equation}
 	\begin{aligned}
 	y_{q,n}[k]=&\widetilde{\mathbf{h}}_{\rm B}[k]\mathbf{V}_q\mathbf{H}_{\rm U}[k]\mathbf{f}_{n}[k]+\widetilde{n}_{q,n}[k] \\
 	=& \mathbf{v}_q{\rm diag}(\widetilde{\mathbf{h}}_{\rm B}[k])\mathbf{H}_{\rm U}[k]\mathbf{f}_n[k]+\widetilde{n}_{q,n}[k].
 	 	\end{aligned}
 \end{equation}
 
 By collecting the training signals in $Q$ subframes on subcarrier $k$, yielding 
 \begin{equation}\label{Yk}
\mathbf{Y}[k]=\widetilde{\mathbf{V}}[k]\mathbf{H}_{\rm U}[k]\mathbf{F}[k]+\widetilde{\mathbf{N}}[k],
 \end{equation}
where $\mathbf{Y}[k]\in\mathbb{C}^{Q\times N_X}$, $\widetilde{\mathbf{V}}[k]\triangleq[\mathbf{v}_1^T,\cdots,\mathbf{v}_Q^T]^T{\rm diag}(\widetilde{\mathbf{h}}_{\rm B}[k])\in\mathbb{C}^{Q\times L}$, $\mathbf{F}[k]\triangleq\left[\mathbf{f}^T_1[k],\cdots, \mathbf{f}^T_{N_X}[k]\right]^T\in\mathbb{C}^{N_T\times N_X}$, and $\widetilde{\mathbf{N}}[k]\in\mathbb{C}^{Q\times N_X}$.
 
 \vspace{-0.33cm}
 \section{Near-Field Dictionary Design With Beam Squint}\label{widebanddic}
 The goal of this section is to estimate $\{\mathbf{H}_{\rm U}[k]\}_{k=1}^K$ from Eqn. (\ref{Yk}). One straightforward method is to use the conventional estimation method for each subcarrier. The 2D least squares (2D-LS) estimator for the $k$-th subcarrier is described in the following proposition.
 \begin{Proposition}\label{2D-LSest}
 Assuming $Q=N_R$ and $N_X=N_U$, the solution of $\mathbf{H}_{\rm U}[k]$ of Eqn. (\ref{Yk}) via the 2D-LS estimator is given by
 \begin{equation}\label{2DLS}
\widehat{\mathbf{H}}_{\rm U}[k]=\left(\widetilde{\mathbf{V}}^H[k]\widetilde{\mathbf{V}}[k]\right)^{-1}\widetilde{\mathbf{V}}^H[k]\mathbf{Y}[k]\mathbf{F}^H[k]\left(\mathbf{F}[k]\mathbf{F}^H[k]\right)^{-1}.
 \end{equation}
 \end{Proposition}
\begin{Proof}
According to \cite[Chapter 2]{2DLSE}, in the case when
\begin{equation}
\left(\widetilde{\mathbf{V}}^H[k]\widetilde{\mathbf{V}}[k]\right)^{-1}\widetilde{\mathbf{V}}^H[k]\widetilde{\mathbf{V}}[k]\mathbf{Y}[k]\mathbf{F}[k]\mathbf{F}^H[k]\left(\mathbf{F}[k]\mathbf{F}^H[k]\right)^{-1}=\mathbf{Y}[k],
\end{equation} 
 the general solution of $\mathbf{H}_{\rm U}[k]$ is 
\begin{equation}
	\begin{aligned}
\widehat{\mathbf{H}}_{\rm U}[k]=&\left(\widetilde{\mathbf{V}}^H[k]\widetilde{\mathbf{V}}[k]\right)^{-1}\widetilde{\mathbf{V}}^H[k]\mathbf{Y}[k]\mathbf{F}^H[k]\left(\mathbf{F}[k]\mathbf{F}^H[k]\right)^{-1}+\mathbf{D}\\
&\  -\left(\widetilde{\mathbf{V}}^H[k]\widetilde{\mathbf{V}}[k]\right)^{-1}\widetilde{\mathbf{V}}^H[k]\widetilde{\mathbf{V}}[k]\mathbf{D}\mathbf{F}[k]\mathbf{F}^H[k]\left(\mathbf{F}[k]\mathbf{F}^H[k]\right)^{-1}
	\end{aligned}
\end{equation}
for aribitrary $\mathbf{D}\in\mathbb{C}^{N_R\times N_U}$.

With the appropriate training sequence design, we have $\left(\widetilde{\mathbf{V}}^H[k]\widetilde{\mathbf{V}}[k]\right)^{-1}\widetilde{\mathbf{V}}^H[k]\widetilde{\mathbf{V}}[k]=\mathbf{I}_{N_R}$ and $\mathbf{F}[k]\mathbf{F}^H[k]\left(\mathbf{F}[k]\mathbf{F}^H[k]\right)^{-1}=\mathbf{I}_{N_U}$. Thus, Eqn. (\ref{2DLS}) holds. 

The proof is complete.
\end{Proof}
\begin{remark}
The channel $\mathbf{H}_{\rm U}[k]$ can also be solved by the 1D-LS estimator with the Kronecker product.
Ignoring the noise and vectorizing $\mathbf{Y}[k]$ to obtain ${\rm vec}(\mathbf{Y}[k])\approx(\mathbf{F}^T[k]\otimes\widetilde{\mathbf{V}}[k]){\rm vec}(\mathbf{H}_{\rm U}[k])$. Then, $\mathbf{H}_{\rm U}[k]$ can be solved by
\begin{equation}\label{1DLS}
\widehat{\mathbf{H}}_{\rm U}[k]={\rm devec}\left(\left(\mathbf{F}_{\rm V}^H[k]\mathbf{F}_{\rm V}[k]\right)^{-1}\mathbf{F}_{\rm V}^H[k]{\rm vec}(\mathbf{Y}[k])\right),
\end{equation} 
where ${\rm devec}(\cdot)$ is the de-vectorization operator, and 
$\mathbf{F}_{\rm V}[k]\triangleq \mathbf{F}^T[k]\otimes\widetilde{\mathbf{V}}[k]\in\mathbb{C}^{LN_U\times LN_U}$.

However, this way will incur much higher computational complexity compared with the 2D-LS estimator (please see details in the later complexity analysis section). 
\end{remark}

Despite its simplicity, the 2D-LS method has three limitations: 1) the pilot overhead must be equal to or greater than the number of elements, resulting in a significant pilot overhead, 2) it is susceptible to noise, and 3) it cannot perform joint channel estimation for all subcarriers for performance enhancement.

In contrast, the CS technique can effectively overcome these limitations for channel estimation. It is well established that the key to CS is sparse representation. Because of the limited number of channel paths, we can represent $\mathbf{H}_{\rm U}[k]$ sparsely using appropriate dictionaries. Suppose that $\mathbf{B}_{\rm R}[k]\in\mathbb{C}^{L\times G_R}$ and $\mathbf{A}_{\rm U}[k]\in\mathbb{C}^{N_U\times G_U}$ are the dictionaries used to describe the spatial parameters of the RIS and user on subcarrier $k$, respectively, where $G_R$ and $G_U$ are the dictionary sizes. Then, $\mathbf{H}_{\rm U}[k]$ can be approximated by
\begin{equation}\label{repre}
\mathbf{H}_{\rm U}[k]\approx \mathbf{B}_{\rm R}[k]\bm{\Xi}[k]\mathbf{A}_{\rm U}^H[k],
\end{equation}  
where  $\bm{\Xi}[k]\in\mathbb{C}^{G_R\times G_U}$ denotes a $P$-sparse matrix.

Note that designing $\mathbf{B}_{\rm R}[k]$ is more challenging than designing $\mathbf{A}_{\rm U}[k]$ because of the near-field effect. To develop a wideband near-field dictionary, we first need to discuss the beam squint effect in the near-field region. Since this paper's following derivation will not consider the recovery of the BS-RIS channel parameter, we will ignore the angle subscript for simplicity, such as $\widetilde{\theta}\leftarrow \widetilde{\theta}_{\rm R}$ and $\widetilde{\vartheta}\leftarrow \widetilde{\vartheta}_{\rm U}$.
  \begin{figure}
 	\centering
 	\includegraphics[width = 0.415\textwidth]{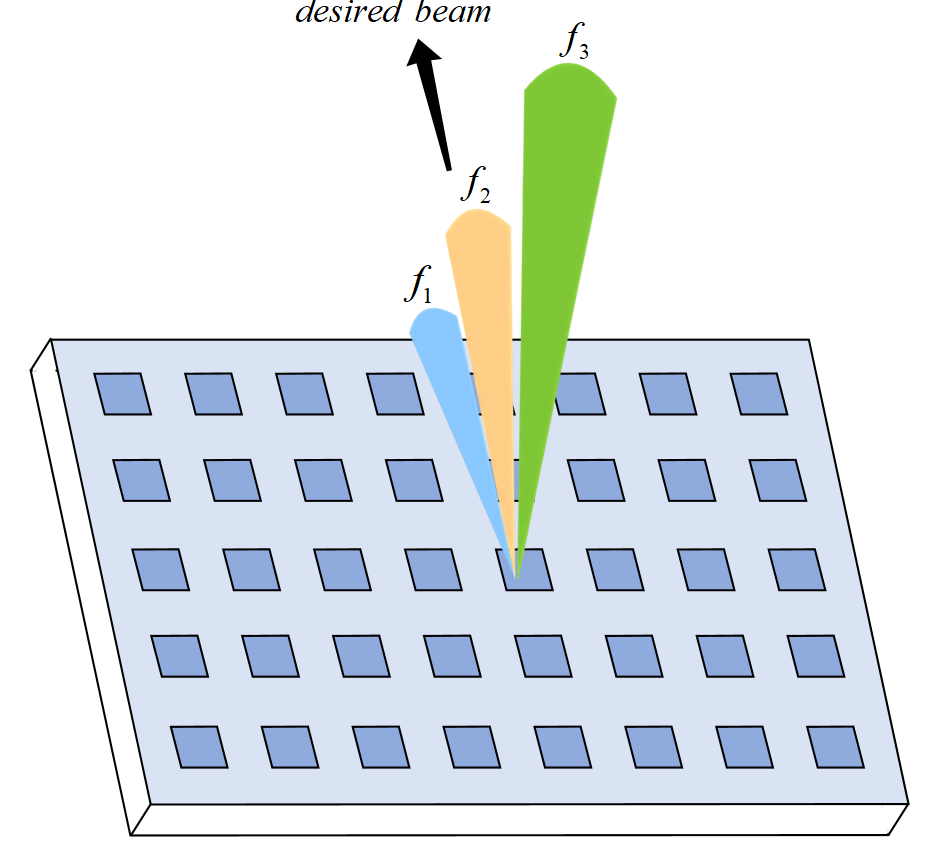}
 	\caption{The near-field beam squint effect.}
 	\label{nbs}
 \end{figure}
\vspace{-0.33cm}
\subsection{Near-Field Spherical-Domain Beam Squint}\label{NF-BS} 
Let the index of the central frequency $f_c$ be $\bar{k}\triangleq\frac{K+1}{2}$ such that $f_{\bar{k}}=f_c$. Given a desired beam on frequency $f_c$,
the near-field effect indicates that the steering beams of other frequencies will have a spatial shift, as shown in Fig. \ref{nbs}. Next, we will derive the mathematical expression to describe the spatial shift.

Reference \cite{near-field1} showed that the Fresnel approximation, with  neglection of terms  greater than quadratic,
was sufficient to characterize near-field communications. According to this, we can simplify the near-field array response.
Defining the phase term
$\delta_{k,(m^y_R,m^z_R)}\triangleq \frac{f_k}{c}( -m^y_Rd \widetilde{\phi} - m^z_Rd\widetilde{\theta} +\frac{(m^z_Rd)^2}{2r}(1-\widetilde{\theta}^2)   +\frac{(m^y_Rd)^2}{2r}(1-\widetilde{\phi}^2)-\frac{m^y_Rm^z_Rd^2\widetilde{\theta}\widetilde{\phi}}{r}) $ by simplifying the phase term of Eqn. (\ref{bB}).
Then, the spherical array response is defined as
\begin{equation}\label{cR}
	\begin{aligned}
		 \mathbf{c}_{\rm R}(f_k,\widetilde{\theta},\widetilde{\phi},r)\triangleq 	 \frac{1}{\sqrt{N_R}}
		\left[ e^{-j{2\pi} \delta_{k,(-M^y_R,-M^z_R)}},\cdots,e^{-j{2\pi} \delta_{k,(M^y_R,M^z_R)}} \right]^T,
	\end{aligned}
\end{equation}

 Suppose that the $\bar{k}$-th subcarrier's phase with a desired beam $(\widetilde{\theta}_{\bar{k}},\widetilde{\phi}_{\bar{k}},{r}_{\bar{k}})$ is defined by $\delta_{\bar{k},(m^y_R,m^z_R)}\triangleq \frac{f_{\bar{k}}}{c}( -m^y_Rd \widetilde{\phi}_{\bar{k}} - m^z_Rd\widetilde{\theta}_{\bar{k}} +\frac{(m^z_Rd)^2}{2r}(1-\widetilde{\theta}_{\bar{k}}^2)   +\frac{(m^y_Rd)^2}{2r}(1-\widetilde{\phi}_{\bar{k}}^2)-\frac{m^y_Rm^z_Rd^2\widetilde{\theta}_{\bar{k}}\widetilde{\phi}_{\bar{k}}}{r} )$. 
  The pattern function pointing to the beam $(\widetilde{\theta}_{\bar{k}},\widetilde{\phi}_{\bar{k}},{r}_{\bar{k}})$ on subcarrier $k$, denoted by  $\bm{\mathcal{S}}(\widetilde{\theta},\widetilde{\phi},r)[k]$ with $ \{\widetilde{\theta},\widetilde{\phi}\}\in[-1,1]$ and $r\in(0,+\infty)$, is expressed as
\begin{equation}\label{S}
	\begin{aligned}
	\bm{\mathcal{S}}(\widetilde{\theta},\widetilde{\phi},r)[k]\triangleq  \frac{1}{N_R}\sum_{m_R^y=-M_R^y}^{M_R^y}\sum_{m_R^z=-M_R^z}^{M_R^z} I_{k,(m_R^y,m_R^z)}e^{-j{2\pi}\left( \delta_{k,(m^y_R,m^z_R)}-\delta_{\bar{k},(m^y_R,m^z_R)}\right)}  \times
\bm{\mathcal{E}}_{(m_R^y,m_R^z)}(\widetilde{\theta},\widetilde{\phi},r)[k], 
\end{aligned}
\end{equation}
where $I_{k,(m_R^y,m_R^z)}$ and $\bm{\mathcal{E}}_{(m_R^y,m_R^z)}[k]$ are the  excitation amplitude and the element pattern of the $(m_R^y,m_R^z)$-th antenna on subcarrier $k$, respectively. As the passive RIS structure is considered, we have $I_{k,(m_R^y,m_R^z)}=1$.
 Moreover, we assume each antenna element on each subcarrier is isotropic such that $\bm{\mathcal{E}}_{(m_R^y,m_R^z)}(\widetilde{\theta},\widetilde{\phi},r)[k]=1$.

To make the pattern independent of the subcarrier index, we have
$
	e^{-j{2\pi}\left( \delta_{k_1,(m^y_R,m^z_R)}-\delta_{\bar{k},(m^y_R,m^z_R)}\right)}=	e^{-j{2\pi}\left( \delta_{k_2,(m^y_R,m^z_R)}-\delta_{\bar{k},(m^y_R,m^z_R)}\right)}
$, where $k_1,k_2\in\{1,\cdots,K\}$, $k_1\neq k_2$. This equation is equivalent to
\begin{equation}
\delta_{k_1,(m^y_R,m^z_R)}=\delta_{k_2,(m^y_R,m^z_R)}+Z,
\end{equation} 
where $Z$ is any integer. By assuming $Z=0$, $k_1=k$, and $k_2=\bar{k}$, we can calculate $\bar{\delta}_k\triangleq\delta_{k,(m^y_R,m^z_R)}-\delta_{\bar{k},(m^y_R,m^z_R)}$ as
	\begin{equation}\label{Dk}
		\begin{aligned}
			\bar{\delta}_k=&
			-\frac{m_R^yd}{c}(f_k\widetilde{\phi}-f_c\widetilde{\phi}_{\bar{k}})- \frac{m_R^zd}{c}(f_k\widetilde{\theta}-f_c\widetilde{\theta}_{\bar{k}})  +\frac{(m^z_Rd)^2}{2c}\left(f_k\frac{1-\widetilde{\theta}^2}{r}-f_c\frac{1-\widetilde{\theta}_{\bar{k}}^2}{r_{\bar{k}}}\right) \\ &+\frac{(m^y_Rd)^2}{2c}\left(f_k\frac{1-\widetilde{\phi}^2}{r}-f_c\frac{1-\widetilde{\phi}_{\bar{k}}^2}{r_{\bar{k}}}\right) -\frac{m_R^ym_R^zd^2}{c}\left(\frac{f_k\widetilde{\theta}\widetilde{\phi}}{r}-\frac{f_c\widetilde{\theta_{\bar{k}}}\widetilde{\phi}_{\bar{k}}}{r_{\bar{k}}}\right). 
		\end{aligned}
	\end{equation}

Now, the problem is equivalent to maximizing $\left\vert\bm{\mathcal{S}}(\widetilde{\theta},\widetilde{\phi},r)[k]\right\vert$ when assigning zero to $\bar{\delta}_k$. In this way,
the method in \cite{RIS-NF-BSE2} used for ULAs is extended to calculate the $k$-the subcarrier's $(\widetilde{\theta},\widetilde{\phi},r)$ that is independent on the antenna index. 
 By assigning zero to $\bar{\delta}_k$, the following system of equations is derived:
\begin{eqnarray}
	\begin{cases}
		f_k\widetilde{\phi}-f_c\widetilde{\phi}_{\bar{k}}=0, \	f_k\widetilde{\theta}-f_c\widetilde{\theta}_{\bar{k}}=0,\\
		f_k\frac{1-\widetilde{\theta}^2}{r}-f_c\frac{1-\widetilde{\theta}_{\bar{k}}^2}{r_{\bar{k}}}=0, \	f_k\frac{1-\widetilde{\phi}^2}{r}-f_c\frac{1-\widetilde{\phi}_{\bar{k}}^2}{r_{\bar{k}}}=0, \\ 
		\frac{f_k\widetilde{\theta}\widetilde{\phi}}{r}-	\frac{f_c\widetilde{\theta}_{\bar{k}}\widetilde{\phi}_{\bar{k}}}{r_{\bar{k}}}=0.
	\end{cases}
\end{eqnarray}

Nevertheless, the system of equations under consideration fails to yield a unique solution due to the interdependent nature of the elevation and azimuth angles on a shared distance parameter, leading to a spatial mismatch. To resolve this issue, we conduct a sequential analysis of each component of the near-field phase.

\subsubsection{Linear Term} 
Since the linear terms dominate the phase in both the near- and far-field region, we first ignore the nonlinear terms to obtain the optimal angles on subcarrier $k$ when maximizing $\left\vert\bm{\mathcal{S}}(\widetilde{\theta},\widetilde{\phi},r)[k]\right\vert$. In this way, we can get $\widetilde{\phi}=\frac{f_c}{f_k}\widetilde{\phi}_{\bar{k}}$ and $\widetilde{\theta}=\frac{f_c}{f_k}\widetilde{\theta}_{\bar{k}}$.

\subsubsection{Quadratic Term}
The quadratic term poses a significant challenge in achieving near-field wideband beam focusing, particularly due to the dependence on the distance parameter, which is further aggravated by the use of UPAs. To address this issue, the authors in \cite[Lemma 3]{Spherdic} have shown that the quadratic cross term $\frac{(m^z_Rd)^2}{2r}(1-\widetilde{\theta}^2)   +\frac{(m^y_Rd)^2}{2r}(1-\widetilde{\phi}^2)-\frac{m^y_Rm^z_Rd^2\widetilde{\theta}\widetilde{\phi}}{r}$ could be neglected to simplify the quadratic form, thus facilitating the design of UPA codebooks. In line with this assumption, we investigate the near-field spherical-domain beam squint effect.
We proceed by focusing on the beam pattern without the linear term and quadratic cross term, i.e., 
\begin{equation}\label{Sk}
	\begin{aligned}
	\left\vert\bm{\mathcal{S}}(\widetilde{\theta},\widetilde{\phi},r)[k]\right\vert=&\frac{1}{N_R}\left\vert\sum_{m_R^y=-M_R^y}^{M_R^y}\sum_{m_R^z=-M_R^z}^{M_R^z}  e^{-j2\pi\left( \frac{(m^z_Rd)^2}{2c}\left(f_k\frac{1-\widetilde{\theta}^2}{r}-f_c\frac{1-\widetilde{\theta}_{\bar{k}}^2}{r_{\bar{k}}}\right)+ \frac{(m^y_Rd)^2}{2c}\left(f_k\frac{1-\widetilde{\phi}^2}{r}-f_c\frac{1-\widetilde{\phi}_{\bar{k}}^2}{r_{\bar{k}}} \right)\right) }\right\vert \\
=&\frac{1}{N_R^y}\left\vert\sum_{m_R^y=-M_R^y}^{M_R^y} e^{j2\pi  \frac{(m^y_Rd)^2}{2c}\left(f_k\frac{1-\widetilde{\phi}^2}{r}-f_c\frac{1-\widetilde{\phi}_{\bar{k}}^2}{r_{\bar{k}}} \right)}\right\vert\times 
\frac{1}{N_R^z} \left\vert \sum_{m_R^z=-M_R^z}^{M_R^z}e^{j2\pi  \frac{(m^z_Rd)^2}{2c}\left(f_k\frac{1-\widetilde{\theta}^2}{r}-f_c\frac{1-\widetilde{\theta}_{\bar{k}}^2}{r_{\bar{k}}}\right)}\right\vert \\
=& g_y(\widetilde{\phi},r,k)g_z(\widetilde{\theta},r,k).
	\end{aligned}
\end{equation}

Since the above product of summation is hard to handle, the following proposition gives the approximation.

\begin{Proposition}\label{proposition1}
 Given the error function ${\rm erf}(x)\triangleq \frac{2}{\sqrt{\pi}}\int_{0}^{x}e^{-t^2}{\rm d}t$, the pattern $\left\vert\bm{\mathcal{S}}(\widetilde{\theta},\widetilde{\phi},r)[k]\right\vert$ without the linear term and the quadratic cross term can be approximated by
\begin{equation}
	\begin{aligned}
\left\vert\bm{\mathcal{S}}(\widetilde{\theta},\widetilde{\phi},r)[k]\right\vert\approx  \widetilde{g}_y(\zeta_{\phi,k})\widetilde{g}_z(\zeta_{\theta,k})
\triangleq  \widetilde{g}(\zeta_{\phi,k},\zeta_{\theta,k})
	\end{aligned}
\end{equation}
where $\widetilde{g}_y(\zeta_{\phi,k})=\left\vert  \frac{1}{\sqrt{\zeta_{\phi,k}}N_R^y}  {\rm erf}\left(
\frac{1-j}{2\sqrt{2}}\sqrt{\pi\zeta_{\phi,k}}N_{R}^y
\right) \right\vert$ and $\widetilde{g}_z(\zeta_{\theta,k})=\left\vert  \frac{1}{\sqrt{\zeta_{\theta,k}}N_R^z}  {\rm erf}\left(
\frac{1-j}{2\sqrt{2}}\sqrt{\pi\zeta_{\theta,k}}N_{R}^z
\right) \right\vert$ with $\zeta_{\phi,k}\triangleq\frac{ d^2}{c}\left(f_k\frac{1-\widetilde{\phi}^2}{r}-f_c\frac{1-\widetilde{\phi}_{\bar{k}}^2}{r_{\bar{k}}} \right)$ and $\zeta_{\theta,k}\triangleq\frac{ d^2}{c}\left(f_k\frac{1-\widetilde{\theta}^2}{r}-f_c\frac{1-\widetilde{\theta}_{\bar{k}}^2}{r_{\bar{k}}} \right)$.
\end{Proposition}
\begin{Proof}
Please see Appendix \ref{TM}.
\end{Proof}
  
  Based on the analysis of the linear terms, we can determine $\widetilde{\phi}\triangleq\frac{f_c}{f_k}\widetilde{\phi}_{\bar{k}}$ and $\widetilde{\theta}\triangleq\frac{f_c}{f_k}\widetilde{\theta}_{\bar{k}}$ to find the optimal value of $r$ that maximizes $\left\vert\bm{\mathcal{S}}(\widetilde{\theta},\widetilde{\phi},r)[k]\right\vert$. However, solving for $r$ analytically becomes challenging due to the multiplication of two error functions. Therefore, we resort to a numerical method to solve for $r$ by defining an inverse function $\widetilde{g}^{-1}(f_k,\widetilde{\theta}_{\bar{k}},\widetilde{\phi}_{\bar{k}},r_{\bar{k}})$ that returns the solution of $\underset{r}{\rm arg \ max} \ \widetilde{g}(\zeta_{\phi,k},\zeta_{\theta,k})$.
 
    \begin{figure}
  	\centering
  	\includegraphics[width = 0.54\textwidth]{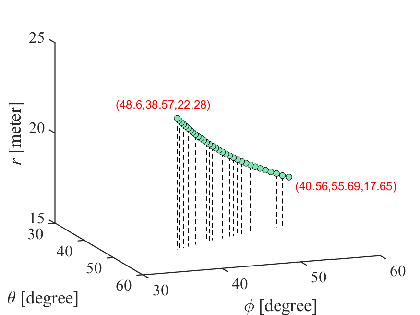}
  	\caption{A trajectory map in the spherical domain regarding $\{\theta,\phi,r,f_k\}$}
  	\label{traj}
  \end{figure}

Let us consider a wideband system with $N_R^y=256$, $N_R^z=4$, $K=32$, and central frequency $f=28$ GHz with a corresponding bandwidth of $f_s=4$ GHz. Assuming that the desired beam parameters for the central frequency are $\theta=\phi=45^\circ$ and $r=20$ meters, a spherical-domain trajectory map with $K$ beams is depicted in Fig. \ref{traj}. Analysis of the figure indicates that the beam point observed at the start and end points, i.e., $(40.56,55.69,17.65)$ and $(48.6,38.57,22.28)$, respectively, reflects the shift of beams at other frequencies when compared to the central frequency at $(45,45,20)$.
\vspace{-0.34cm}
\subsection{Wideband Spherical-Domain Dictionary Design}\label{WD}
For channel estimation using the CS technique, the virtual channel representation with dictionaries is a critical component.
  One commonly used method for designing an effective dictionary $\mathbf{A}\in\mathbb{C}^{N\times G}$ is to minimize the total coherence $\mu=\sum_{g_1}^G\sum_{g_2,g_2\neq g_1}^G\left\vert[\mathbf{A}]_{:,g_1}^H[\mathbf{A}]_{:,g_2}\right\vert$, which is equivalent to minimizing $\left\vert[\mathbf{A}]_{:,g_1}^H[\mathbf{A}]_{:,g_2}\right\vert$ for $\forall g_1,g_2$.
To design the dictionaries $\mathbf{A}_{\rm U}[k]$ and $\mathbf{B}_{\rm R}[k]$, we first derive them with a narrowband system at frequency $f_c$ and then obtain their wideband form.

It is well established that the 2D discrete Fourier transform (2D-DFT) matrix can be adopted to design the 2D angle sampling for dictionary $\mathbf{A}_{\rm U}[\bar{k}]$, which is given by
\begin{equation}\label{AU}
	\begin{aligned}
	\mathbf{A}_{\rm U}[{\bar{k}}]=&\left\{ \mathbf{a}_{\rm U}(f_c,\widetilde{\vartheta},\widetilde{\phi})|\widetilde{\vartheta}=-1,\frac{-G_{ U}^y+2}{G_{ U}^y},\cdots,\frac{G_{ U}^y-2}{G_{ U}^y},  \widetilde{\phi}=-1,\frac{-G_{ U}^z+2}{G_{ U}^z},\cdots,\frac{G_{ U}^z-2}{G_{ U}^z}
	\right\},
		\end{aligned}
\end{equation}
where $G_{ U}^yG_{ U}^z=G_{ U}$ is the total number of atoms.

We aim to orthogonalize two arbitrary atoms (with indices $g_1$ and $g_2$, where $g_1 \neq g_2$) in $\mathbf{B}_{\rm R}[\bar{k}]$ by minimizing the coherence, $\mu_{g_1,g_2}$, approximating it to $0$, where
\begin{equation}
	\mu_{g_1,g_2}=\left\vert\mathbf{c}^H_{\rm R}(f_c,\widetilde{\theta}_{g_1},\widetilde{\phi}_{g_1},r_{g_1})\mathbf{c}_{\rm R}(f_c,\widetilde{\theta}_{g_2},\widetilde{\phi}_{g_2},r_{g_2})\right\vert.
\end{equation}
This method is similar to the one described in Eqn. (\ref{S}), with the difference being that Section \ref{NF-BS} focuses on exploring the maximal pattern gain on different subcarriers, while Section \ref{WD} discusses the minimal coherence of arbitrary atoms in a dictionary independent of the subcarrier index.

As the angle and distance parameters are coupled in $\mu_{g_1,g_2}$, we separate the sampling of angle and distance, as done in \cite{near-CE2}. By considering only the linear term of $\mu_{g_1,g_2}$ and assigning $d=\frac{c}{2f_c}$, it can be expressed as
\begin{equation}
	\begin{aligned}
\mu_{g_1,g_2}= 
\frac{1}{N_R}\left\vert \sum_{m_R^y=-M_R^y}^{M_R^y}  \sum_{m_R^z=-M_R^z}^{M_R^z} e^{j\pi \left( m^y_R (\widetilde{\phi}_{g_2}- \widetilde{\phi}_{g_1})+m^z_R (\widetilde{\theta}_{g_2}- \widetilde{\theta}_{g_1})\right) }
\right\vert.
	\end{aligned}
\end{equation}
In this case, the method for sampling angles is the same as that described in Eqn. (\ref{AU}), which is designed using the 2D-DFT matrix.

After sampling the angle space, we focus on sampling the distance of each angle point. 
Since the angle is fixed ($\widetilde{\theta}_{g_1}=\widetilde{\theta}_{g_2}=\widetilde{\theta}$ and $\widetilde{\phi}_{g_1}=\widetilde{\phi}_{g_2}=\widetilde{\phi}$), the distance sampling becomes solvable. The coherence, $\mu_{g_1,g_2}$, can now be re-written as
\begin{equation}
	\begin{aligned}
		\mu_{g_1,g_2}=\frac{1}{N_R}  \left\vert\sum_{m_R^y=-M_R^y}^{M_R^y}\sum_{m_R^z=-M_R^z}^{M_R^z}  e^{-j\frac{c\pi }{4f_c}\left( {(m^z_R)^2(1-\widetilde{\theta}^2)}+{(m^y_R)^2(1-\widetilde{\phi}^2)}\right)\left(\frac{1}{r_{g_2}}-\frac{1}{r_{g_1}}\right)   }\right\vert .
	\end{aligned}
\end{equation}
It is important to note that the above equation is a special case of Eqn. (\ref{Sk}), where $f_k=f_c$, $\widetilde{\theta}= \widetilde{\theta}_{\bar{k}}$ and $\widetilde{\phi}= \widetilde{\phi}_{\bar{k}}$. This can be directly solved by Prop. \ref{proposition1}. Then, we have\footnote{This can also be derived with the Fresnel integral in reference \cite{Spherdic}.}
\begin{equation}
	\mu_{g_1,g_2}\approx \widetilde{g}(\overline{\zeta}_{\phi},\overline{\zeta}_{\theta}),
\end{equation}
where $\overline{\zeta}_{\phi}\triangleq\frac{ c}{4f_c}\left({1-\widetilde{\phi}^2}\right)\left(\frac{1}{r_{g_2}}-\frac{1}{r_{g_1}}\right)$, $\overline{\zeta}_{\theta}\triangleq\frac{ c}{4f_c}\left({1-\widetilde{\theta}^2}\right)\left(\frac{1}{r_{g_2}}-\frac{1}{r_{g_1}}\right)$, and $\widetilde{g}(\cdot,\cdot)$ is defined in Prop. \ref{proposition1}.

Then, we aim at minimizing $\mu_{g_1,g_2}$ with respect to $\frac{1}{r_{g_2}}-\frac{1}{r_{g_1}}$.
However, obtaining a closed solution for distance sampling is challenging due to the product of two error functions.
Fortunately,
it is possible to sample $\frac{1}{r_{g_1}}$ and $\frac{1}{r_{g_2}}$ with a linear approach based on the numerical solution. Let us define a function $e(\frac{1}{r_{g_2}}-\frac{1}{r_{g_1}})\triangleq\vert\widetilde{g}(\overline{\zeta}_{\phi},\overline{\zeta}_{\theta})\vert$ with respect to distance sampling.
 Assuming that $r_{g_2}<r_{g_1}$, it can be established that $e(\frac{1}{r_{g_2}}-\frac{1}{r_{g_1}})$ decreases as $\frac{1}{r_{g_2}}-\frac{1}{r_{g_1}}$ increases from the general trend since the function $e(\cdot)$ has a form of  $\frac{{\rm erf}(a\sqrt{x}){\rm erf}(b\sqrt{x})}{cx}$.

 Defining an inverse function ${e}^{-1}(\mu_m,\widetilde{\theta},\widetilde{\phi})$ that returns the solution of  
\begin{equation}\label{mum}
 \underset{ \frac{1}{r_{g_2}}-\frac{1}{r_{g_1}}}{\rm arg \ max}\ {e}\left( \frac{1}{r_{g_2}}-\frac{1}{r_{g_1}}\right)=\mu_m,
\end{equation} 
 with the input variables $\widetilde{\theta}$, $\widetilde{\phi}$, and the minimal acceptable coherence $\mu_m$.  This indicates that reciprocal distance can be linearly sampled.
  
  To construct the narrowband dictionary $\mathbf{B}_{\rm R}[\bar{k}]$, the following procedure is designed.  First, determining the angle set $\{\widetilde{\theta}_{\bar{k},i_z},\widetilde{\phi}_{\bar{k},i_y}\}_{i_z=1,i_y=1}^{G_R^z,G_R^y}$ with $\widetilde{\theta}_{\bar{k},i_z}=-1+\frac{2}{G_{\rm R}^z}$ and $\widetilde{\phi}_{\bar{k},i_y}=-1+\frac{2}{G_{\rm R}^y}$. Then, for each angle direction $\{\widetilde{\theta}_{\bar{k},i_z},\widetilde{\phi}_{\bar{k},i_y}\}$, the distance parameter $\{r_{\bar{k},i_y,i_z,i_r}\}_{i_r=1}^{G_R^r}$ can be obtained by $\frac{1}{r_{\bar{k},i_y,i_z,i_r}}=\frac{1}{r_{\rm min}}-i_r{e}^{-1}(\mu_m,\widetilde{\theta}_{\bar{k},i_z},\widetilde{\phi}_{\bar{k},i_y})$, where $r_{\rm min}$ is the allowed minimal communication distance, and $\frac{1}{r_{\bar{k},i_y,i_z,i_r}}> 0$. Finally, $\mathbf{B}_{\rm R}[\bar{k}]\in\mathbb{C}^{N_R\times G_R}$ can be obtained with the response vector of Eqn. (\ref{cR})
, where $G_R$ represents the number of all atoms.

\begin{algorithm}
	\caption{Wideband Spherical-Domain Dictionary Design} 
	\label{AL1}
	\KwIn {The size of array apertures $\{N_R^z,N_R^y\}$, the size of sampling points $\{G_R^z,G_R^y\}$, and the number of subcarriers $K$.}
	\KwOut {Wideband spherical-domain dictionary $\{\mathbf{B}_{\rm R}[k]\}_{k=1}^K$. }
	\Begin{ 
		\For{$k=1,2,\cdots,K$}{
			\For{$i_z=1,2,\cdots,G_{\rm R}^z$}{
				$\widetilde{\theta}_{\bar{k},i_z}=  -1+\frac{2i_z}{G_{\rm R}^z} $.
				\\
				\For{$i_y=1,2,\cdots,G_{\rm R}^z$}{		$\widetilde{\phi}_{\bar{k},i_y}=  -1+\frac{2i_y}{G_{\rm R}^y} $;\\
					$r_{\bar{k},i_y,i_z,i_r}=\frac{1}{r}-{e}^{-1}(\mu_m,\widetilde{\theta}_{\bar{k},i_z},{\widetilde{\phi}_{\bar{k},i_y}})$.\\
					\If{$\frac{1}{r_{\bar{k},i_y,i_z,i_r}}>0$}{ 
					$\mathbf{B}_{\rm R}[k]\leftarrow\mathbf{B}_{\rm R}[k]\cup\mathbf{c}_{\rm R}(f_k,\widetilde{\theta}_{\bar{k},i_z},\widetilde{\phi}_{\bar{k},i_y},r_{\bar{k},i_y,i_z,i_r})$.}
				\Else{break.}
			}}
		}
		
	}
	\Return{$\{\mathbf{B}_{\rm R}[k]\}_{k=1}^K$}
\end{algorithm}
In this context, the development of the wideband dictionary can be achieved by utilizing the response vector with various frequencies.
Basically, the wideband spherical-domain dictionary is developed according to Algorithm \ref{AL1}. Although the inverse function necessitates numerical computation, the dictionary can be generated offline and subsequently utilized for online channel estimation.

 \vspace{-0.43cm}
\section{Near- and Far-Field Channel Parameter Recovery}\label{NCE}

Leveraging on the designed dictionary, the channel parameters can be recovered using the CS theory.
Substituting the sparse channel representation (\ref{repre}) into the signal model (\ref{Yk}) results in
 \begin{equation}\label{Ykk}
 	\begin{aligned}
	\mathbf{Y}[k]=&\widetilde{\mathbf{V}}[k]\mathbf{H}_{\rm U}[k]\mathbf{F}[k]+\widetilde{\mathbf{N}}[k] \\
	\approx &\widetilde{\mathbf{V}}[k]\mathbf{B}_{\rm R}[k]\bm{\Xi}[k]\mathbf{A}_{\rm U}^H[k]\mathbf{F}[k]+\widetilde{\mathbf{N}}[k].
	\end{aligned}
\end{equation}

One straightforward approach to  recovering the parameter set $\{\phi,\theta,r,\varphi,\vartheta\}$ is the KCS framework, which converts the multi-dimensional recovery problem into a 1D recovery problem.
Vectoring $\mathbf{Y}[k]$ and ignoring the noise yield
 \begin{equation}
 	\begin{aligned}
 {\rm vec}(\mathbf{Y}[k])\approx \left((\mathbf{A}_{\rm U}^H[k]\mathbf{F}[k])^T\otimes \widetilde{\mathbf{V}}[k]\mathbf{B}_{\rm R}[k]\right){\rm vec}(\bm{\Xi}[k]) .
 	\end{aligned}
 \end{equation}
By denoting $ \mathbf{\mathring{y}}[k]\triangleq{\rm vec}(\mathbf{Y}[k])\in\mathbb{C}^{QN_X\times 1}$, $\bm{\mathring{\xi}}[k]\triangleq{\rm vec}(\bm{\Xi}[k])\in\mathbb{C}^{G_UG_R\times 1}$, and $\bm{\mathring{\Phi}}[k]\triangleq(\mathbf{A}_{\rm U}^H[k]\mathbf{F}[k])^T\otimes \left( \widetilde{\mathbf{V}}[k]\mathbf{B}_{\rm R}[k]\right)\in\mathbb{C}^{QN_X\times G_UG_R}$, the following CS recovery problem is formulated for $\forall k$:
\begin{equation}\label{KCS}
\begin{aligned}
 \mathbf{(P1)} & \ \  \  \underset{\bm{\mathring{\xi}}[k]}{\rm arg \ min}\left\Vert \bm{\mathring{\xi}}[k]\right\Vert_0 \\
&  {\rm s.t.} \ \left\Vert\mathbf{\mathring{y}}[k]-\bm{\mathring{\Phi}}[k]\bm{\mathring{\xi}}[k]\right\Vert_2^2\le \epsilon,
\end{aligned} 
\end{equation}
 where $\epsilon$ is the precise factor.

Although the channel parameter can be recovered in this way, it is computationally intensive since the Kronecker product generates a lengthy vector and large-scale dictionary. In XL systems, this method is too computationally expensive to use.
Moreover, it is challenging to get high-resolution data with this framework due to the coupling of several parameters.

In order to avoid globally processing the lengthy signal and the Kronecker dictionary, we develop the MMPSR method, for fast and effective recovery of  $\{\bm{\Xi}[k]\}_{k=1}^K$. 
Denoting $\bm{\Phi}_{\rm R}[k]\triangleq\widetilde{\mathbf{V}}[k]\mathbf{B}_{\rm R}[k]$ and $\bm{\Phi}_{\rm U}[k]\triangleq\mathbf{F}^H[k]\mathbf{A}_{\rm U}[k]$, Eqn. (\ref{Yk}) is re-written as
\begin{equation}\label{Ykkk}
	\mathbf{Y}[k]\approx\bm{\Phi}_{\rm R}[k] \bm{\Xi}[k] \bm{\Phi}^H_{\rm U}[k]+\widetilde{\mathbf{N}}[k].
\end{equation}

Note that each atom within the wideband spherical dictionary across all subcarriers is generated using a common channel parameter, hence referred to as the common support, i.e.,
\begin{equation}\label{supp}
{\rm supp}(\bm{\Xi}[k_1])={\rm supp}(\bm{\Xi}[k_2]), \ \forall k_1,k_2=1,\cdots,K, \ k_1\neq k_2,
\end{equation}
where $\rm supp(\cdot)$ denotes the sparsity support which contains the non-zero index of a sparse vector/matrix.

To jointly recover the channel parameter with the common sparsity support on all subcarriers, the multi-frequency recovery problem is formulated as
\begin{equation}\label{PSR}
	\begin{aligned}
	&\mathbf{(P2)}    \ \ \ \ \underset{\bm{ {\Xi}}[k]}{\rm arg \ min}\sum_{k=1}^{K}\left\Vert \bm{ {\Xi}}[k]\right\Vert_0 \\
		& \ \ \ \ \
		{\rm s.t.} \ 	{\rm supp}(\bm{ {\Xi}}[1])=\cdots={\rm supp}(\bm{ {\Xi}}[K]) ,\\
		& \ \ \ \ \ \
		\sum_{k=1}^{K}\left\Vert\mathbf{ {Y}}[k]-\bm{\Phi}_{\rm R}[k] \bm{\Xi}[k] \bm{\Phi}^H_{\rm U}[k]\right\Vert_2^2\le \epsilon.
	\end{aligned} 
\end{equation}
This problem bears resemblance to the multiple measurement vector problem \cite{MMV}, which capitalizes on the sparsity of various snapshots to retrieve the signal through a common sensing matrix. However, the distinguishing factor is that the sensing matrix $\bm{\mathring{\Phi}}[k]$ is reliant on the subcarrier index.

To resolve problem (P2), one can identify the atom support and subsequently obtain an estimated channel using the 2D-LS estimator. The following proposition is formulated to reconstruct the channel through the utilization of the atom support.

\begin{Proposition}\label{LSE}
	Given the estimated channel supports $\widehat{\mathbf{B}}_{\rm R}[k]$ and $\widehat{\mathbf{A}}_{\rm U}[k]$, the RIS-user channel can be reconstructed by  
\begin{equation}
	\widehat{\mathbf{H}}_{\rm U}[k]=\widehat{\mathbf{B}}_{{\rm R}}[k]\left(\widehat{\bm{\Phi}}_{{\rm R}}^H[k]\widehat{\bm{\Phi}}_{{\rm R}}[k]\right)^{-1}\widehat{\bm{\Phi}}_{{\rm R}}^H[k]\mathbf{Y}[k]\widehat{\bm{\Phi}}_{{\rm U}}[k]\left(\widehat{\bm{\Phi}}_{{\rm U}}^H[k]\widehat{\bm{\Phi}}_{{\rm U}}[k]\right)^{-1}\widehat{\mathbf{A}}_{{\rm U}}^H[k],
\end{equation} 
where $\widehat{\bm{\Phi}}_{\rm R}[k]\triangleq\widetilde{\mathbf{V}}[k]\widehat{\mathbf{B}}_{\rm R}[k]$ and $\widehat{\bm{\Phi}}_{\rm U}[k]\triangleq\mathbf{F}^H[k]\widehat{\mathbf{A}}_{\rm U}[k]$.
\end{Proposition}

\begin{Proof}
 With the estimator channel support, Eqn. (\ref{Ykkk}) can be expressed as 
 \begin{equation}
\mathbf{Y}[k]\approx\widehat{\bm{\Phi}}_{\rm R}[k] \widehat{\bm{\Xi}}[k] \widehat{{\bm{\Phi}}}^H_{\rm U}[k]+\widetilde{\mathbf{N}}[k],
 \end{equation} 
Via the 2D-LS estimator in Prop. \ref{2D-LSest}, $\widehat{\bm{\Xi}}[k]$ is then got by
\begin{equation}
	\widehat{\bm{\Xi}}[k]=\left(\widehat{\bm{\Phi}}_{{\rm R}}^H[k]\widehat{\bm{\Phi}}_{{\rm R}}[k]\right)^{-1}\widehat{\bm{\Phi}}_{{\rm R}}^H[k]\mathbf{Y}[k]\widehat{\bm{\Phi}}_{{\rm U}}[k]\left(\widehat{\bm{\Phi}}_{{\rm U}}^H[k]\widehat{\bm{\Phi}}_{{\rm U}}[k]\right)^{-1}.
\end{equation} 
Thus, the estimated channel can be written as
\begin{equation}
	\widehat{\mathbf{H}}_{\rm U}[k]=\widehat{\mathbf{B}}_{{\rm R} }[k]\widehat{\bm{\Xi}}[k]\widehat{\mathbf{A}}_{{\rm U} }^H[k].
\end{equation}

The proof is complete.
\end{Proof}

By relying on the insight provided by Prop. \ref{LSE}, the task of determining the channel support, which corresponds to recovering the channel parameters, can be accomplished without resorting to the Kronecker product. This is achievable by taking into account the 2D structure of $\bm{\Xi}[k]$. Additionally, it is beneficial to leverage the common sparsity support exhibited in problem (P2) to improve channel estimation. In light of this, we draw inspiration from the parallelizable CS framework \cite{SVD} and put forth Prop. \ref{svd} to address the task of solving for the common channel support.

\begin{Proposition}\label{svd}
Consider singular value decomposition (SVD) for $\mathbf{Y}[k]=\mathbf{T}_{\rm R}[k]\bm{\Sigma}[k]\mathbf{T}_{\rm U}^H[k]\approx\overline{\mathbf{T}}_{\rm R}[k]\overline{\bm{\Sigma}}[k]\overline{\mathbf{T}}_{\rm U}[k]$, where $\overline{\mathbf{T}}_{\rm R}[k]\in\mathbb{C}^{Q\times P}$, $\overline{\bm{\Sigma}}[k]\in\mathbb{C}^{P\times P}$, and $\overline{\mathbf{T}}_{\rm U}[k]\in\mathbb{C}^{P\times N_X}$ are the decomposed matrices corresponding to the largest $P$ singular values. Then,
the atom supports $\Lambda_{\rm R}$ and $\Lambda_{\rm U}$ are $\{{\rm supp}(\bm{\xi}_{{\rm R},p}[k])\}_{p=1}^P$ and $\{{\rm supp}(\bm{\xi}_{{\rm U},p}[k])\}_{p=1}^P$, respectively,
where $\bm{\xi}_{{\rm R},p}[k]\in\mathbb{C}^{G_R\times 1}$ and $\bm{\xi}_{{\rm U},p}[k]\in\mathbb{C}^{G_U\times 1}$, for $\forall p$, are the 1-sparse vectors, which can be
 obtained by solving the following problem.
\begin{equation}\label{svd_psr}
	\begin{aligned}
		&\mathbf{(P3)}   \ \ \ \  \underset{\bm{{\xi}}_{i,p}[k]}{\rm arg \ min}\sum_{k=1}^{K}\left\Vert \bm{{\xi}}_{i,p}[k]\right\Vert_0 \\
		& \ \ \	{\rm s.t.} \
		{\rm supp}(\bm{{\xi}}_{i,p}[1])=\cdots={\rm supp}(\bm{{\xi}}_{i,p}[K]) ,\\
		& \ \ \ \ \ \ \sum_{k=1}^{K}\left\Vert\widetilde{\mathbf{{t}}}_{i,p}[k]- \bm{ {\Phi}}_i[k]\bm{{\xi}}_{i,p}[k]\right\Vert_2^2\le \epsilon,
	\end{aligned} 
\end{equation}  
where the subscript $i\in\{\rm R,U\}$, $\widetilde{\mathbf{t}}_{{\rm R},p}[k]$ and $\widetilde{\mathbf{t}}_{{\rm U},p}[k]$
represent the $p$-th column of $\overline{\mathbf{T}}_{\rm R}[k](\overline{\bm{\Sigma}}[k])^{\frac{1}{2}}$  and $\overline{\mathbf{T}}_{\rm U}[k](\overline{\bm{\Sigma}}^H[k])^{\frac{1}{2}}$, respectively.

\end{Proposition}

\begin{Proof}
Please see Appendix \ref{appendixB}.
\end{Proof}

\begin{algorithm}
	\caption{Multi-Measurement Parallelizable Subspace Recovery} 
	\label{AL2}
	\KwIn {The received signal $\mathbf{Y}[k]$, measurement matrices $\widetilde{\mathbf{V}}[k]$ and $\mathbf{F}[k]$, dictionaries $\mathbf{B}_{\rm R}[k]$ and $\mathbf{A}_{\rm U}[k]$
		, $\forall k =1,\cdots, K$.
	}
	\KwOut {Estimated channels $\{\mathbf{H}_{\rm U}[k]\}_{k=1}^K$.}
	\Begin{
		Construct $\bm{\Phi}_{\rm R}[k]\triangleq\widetilde{\mathbf{V}}[k]\mathbf{B}_{\rm R}[k]$ and $\bm{\Phi}_{\rm U}[k]\triangleq\mathbf{F}^H[k]\mathbf{A}_{\rm U}[k]$. \\  Obtain $\mathbf{Y}[k]\approx\overline{\mathbf{T}}_{\rm R}[k]\overline{\bm{\Sigma}}[k]\overline{\mathbf{T}}_{\rm U}[k]$ via SVD. \\
		\For{$p=1,2,\cdots,P$}{
			
			\For{ $k=1,2,\cdots,K$  }{Calculate the CC vectors $\mathring{\bm{\rho}}_{{\rm R},p}[k]$ and $\mathring{\bm{\rho}}_{{\rm U},p}[k]$ based on Eqn. (\ref{CC}).    
			}
			$\mathring{g}_{r,p}=\underset{{g}_r}{\rm arg \ max} \ \sum_{k=1}^K \left\vert \left[\mathring{\bm{\rho}}_{{\rm R},p}[k]\right]_{g_r}\right\vert$,  $\mathring{g}_{u,p}=\underset{{g}_u}{\rm arg \ max} \ \sum_{k=1}^K\left\vert \left[\mathring{\bm{\rho}}_{{\rm U},p}[k]\right]_{g_u}\right\vert$. \\
			Obtain the estimated parameters $\{\mathring{\theta}_p,\mathring{\phi}_p,\mathring{r}_p,\mathring{\vartheta}_p,\mathring{\varphi}_p\}$ according to $\mathring{g}_{r,p}$ and $\mathring{g}_{u,p}$.\\
			\For{ $k=1,2,\cdots,K$  }{Calculate the refined CC vectors $\breve{\bm{\rho}}_{{\rm R},p}[k]$ and $\breve{\bm{\rho}}_{{\rm U},p}[k]$ based on Eqn. (\ref{CC}).    
			}  
			$\breve{g}_{r,p}=\underset{{g}_r}{\rm arg \ max} \ \sum_{k=1}^K\vert \left[\breve{\bm{\rho}}_{{\rm R},p}[k]\right]_{g_r}\vert$,  $\breve{g}_{u,p}=\underset{{g}_u}{\rm arg \ max} \ \sum_{k=1}^K\vert \left[\breve{\bm{\rho}}_{{\rm U},p}[k]\right]_{g_u}\vert$.\\
			Obtain the refined parameters $\{\breve{\theta}_p,\breve{\phi}_p,\breve{r}_p,\breve{\vartheta}_p,\breve{\varphi}_p\}$ according to $\breve{g}_{r,p}$ and $\breve{g}_{u,p}$.
		}
		\For{$k=1,2,\cdots,K$}{	Reconstruct the channel $\mathbf{H}_{\rm U}[k]$ according to the refined parameters and Prop. \ref{LSE}.
	}}
	\Return{$\{\mathbf{H}_{\rm U}[k]\}_{k=1}^K$}  
\end{algorithm}
\begin{remark}
This framework enables the decoupling of the high-dimensional recovery problem into a series of simple subproblems that can be solved in parallel. Additionally, the size of the dictionary for each subproblem is small, which reduces complexity and facilitates super-resolution or refined procedures.
\end{remark}

Based on Prop. \ref{svd} and Prop. \ref{LSE}, problem (P2) is simplified into problem (P3), which can be solved by the following steps. 
\subsubsection{CC-Based Atom Matching} 
Atom matching is an effective solution for problem (P3) due to the $1$-sparsity of $\bm{{\xi}}_{i,p}[k]$. In the greedy recovery algorithms (such as OMP and subspace pursuit), atom matching is performed by taking the inner product of the measured vector and the atom. Since the inner product depends not only on the phase but also on the modulus, the matching result can be influenced by the modulus of the atom, which varies for each atom in the sensing matrix. However, we are only concerned with the accuracy of the matched phase. To address this, we use CC which relies only on the phase difference between two vectors to perform atom matching.
 
Consider the two arbitrary vectors $\mathbf{x}\in\mathbb{C}^{N\times 1}$ and $\mathbf{y}\in\mathbb{C}^{N\times 1}$, the CC is given by
\begin{equation}\label{CC}
	\rho=\frac{\sum_{n=1}^{N}(\mathbf{x}_n-{\bar{\mathbf{x}}_n})(\mathbf{y}_n-{\bar{\mathbf{y}}_n})}{\sqrt{\sum_{n=1}^{N}(\mathbf{x}_n-{\bar{\mathbf{x}}_n})^2}\sqrt{\sum_{n=1}^{N}(\mathbf{y}_n-{\bar{\mathbf{y}}_n})^2}},
\end{equation}
where $\mathbf{x}_n\triangleq [\mathbf{x}]_n$, $\mathbf{y}_n\triangleq [\mathbf{y}]_n$, $\bar{\mathbf{x}}_n\triangleq\frac{1}{N}\sum_{n=1}^{N}[\mathbf{x}]_n$, and $\bar{\mathbf{y}}_n\triangleq\frac{1}{N}\sum_{n=1}^{N}[\mathbf{y}]_n$.

According to Eqn. (\ref{CC}), we define the CC vector $\mathring{\bm{\rho}}_{i,p}[k]\in\mathbb{C}^{G_i\times 1}$ ($i\in\{{\rm R,U}\},p\in\{1,\cdots,P\}$), which consists of the CCs between $\widetilde{\mathbf{t}}_{i,p}[k]$ and the columns of $\bm{\Phi}_i[k]$. 
\subsubsection{Joint Multi-Frequency Processing}
The vector $\mathring{\bm{\rho}}_{i,p}[k]$ specifies the atom to be used to represent $\widetilde{\mathbf{t}}_{i,p}[k]$. 
 According to Eqn. (\ref{supp}), the best atom for $\forall \{i,p\}$ can be selected across all subcarriers. Defining $\mathring{g}_{r,p}$ and $\mathring{g}_{u,p}$ as the indices of the best atoms $\bm{\xi}_{{\rm R},p}[k]$ and $\bm{\xi}_{{\rm U},p}[k]$, $\forall k$, respectively, and we have 
$\mathring{g}_{r,p}=\underset{{g}_r}{\rm arg \ max} \ \sum_{k=1}^K\vert \left[\mathring{\bm{\rho}}_{{\rm R},p}[k]\right]_{g_r}\vert$,  $\mathring{g}_{u,p}=\underset{{g}_u}{\rm arg \ max} \ \sum_{k=1}^K\vert \left[\mathring{\bm{\rho}}_{{\rm U},p}[k]\right]_{g_u}\vert$. 
\subsubsection{Parameter Refinement}
It should be noted that the estimation error depends on the resolution of the dictionary. However, in order to avoid high computing costs and reduce storage burden, the dictionary is always low-resolution, which may lead to a high channel estimation error, particularly in multi-parameter recovery problems. Therefore, designing a refinement procedure for the atom support is essential for improving estimation performance.
 
Let us consider angle refinement first. The angles that correspond to the estimated support are defined by $\{\mathring{\theta}_p,\mathring{\phi}_p,\mathring{\vartheta}_p,\mathring{\varphi}_p\}$ for $\forall p$. Note that the estimated angles are distributed between the two sampling points of their respective dictionaries.
Refined angles $\{\breve{\theta}_p,\breve{\phi}_p,\breve{\vartheta}_p,\breve{\varphi}_p\}$ can be obtained by sampling uniformly with angle stepsizes  $\{\Delta_\theta,\Delta_\phi,\Delta_\vartheta,\Delta_\varphi\}$, such as
$\breve{\theta}\in\left\{\widetilde{\theta} \Big|\widetilde{\theta}=\mathring{\theta}-\frac{1}{G_R^z}:\Delta_\theta:\mathring{\theta}+\frac{1}{G_R^z}\right\}$ and
$\breve{\vartheta}\in\left\{\widetilde{\vartheta} \Big|\widetilde{\vartheta}=\mathring{\vartheta}-\frac{1}{G_U^y}:\Delta_\vartheta:\mathring{\vartheta}+\frac{1}{G_U^y}\right\}$. 
 Likewise, the refined distance $\breve{r}$ can be derived from a new sample point set with the distance stepsize $\Delta_r$: $\frac{1}{\breve{r}}\in\left\{ \frac{1}{r} \Big| \frac{1}{r}= \frac{1}{\mathring{r}}-\frac{{e}^{-1}(\mu_m,\mathring{{\theta}},{\mathring{\phi}})}{2}:\Delta_r:\frac{1}{\mathring{r}}+\frac{{e}^{-1}(\mu_m,\mathring{{\theta}},{\mathring{\phi}})}{2}
 \right\}$.
 With the new sampling points, parameters can be refined by atom matching and joint multi-frequency processing as described above. Specifically, the CC vector between the refined spherical-domain support generated by the above parameters and $\widetilde{\mathbf{t}}_{{\rm R},p}[k]$ is denoted by $\breve{\bm{\rho}}_{{\rm R},p}[k]\in\mathbb{C}^{\frac{4{e}^{-1}(\mu_m,\mathring{{\theta}},{\mathring{\phi}})}{G_{\rm R}^zG_{\rm R}^y\Delta_{\rm R}}\times 1}$ for $\forall k$, where $\Delta_{\rm R}\triangleq\Delta_\theta\Delta_\phi\Delta_r$. Similarly, the CC vector regarding the refined angular-domain support is denoted by $\breve{\bm{\rho}}_{{\rm U},p}\in\mathbb{C}^{\frac{4}{G_{\rm U}\Delta_{\rm U}}\times 1}$ for $\forall k$, where $\Delta_{\rm U}=\Delta_\vartheta\Delta_\varphi$. Then, the best indices $\{\breve{g}_{r,p},\breve{g}_{u,p}\}$ can be found by
 $\breve{g}_{r,p}=\underset{{g}_r}{\rm arg \ max} \ \sum_{k=1}^K\vert \left[\breve{\bm{\rho}}_{{\rm R},p}[k]\right]_{g_r}\vert$,  $\breve{g}_{u,p}=\underset{{g}_u}{\rm arg \ max} \ \sum_{k=1}^K\vert \left[\breve{\bm{\rho}}_{{\rm U},p}[k]\right]_{g_u}\vert$. 
\vspace{-0.5cm}
\section{Lower Bound Derivation}\label{LBD}

Based on the oracle least squares (OLS) estimator \cite{OLS1,OLS2} that serves as the lower bound with the perfectly known channel support, this paper proposes a 2D-OLS  estimator for 2D-CS and derives its  theoretical NMSE bound.

Recalling Eqn. (\ref{H_RU}), the channel is re-written by a compact form:
 $\mathbf{H}_{\rm U}[k]\triangleq\overline{\mathbf{B}}_{\rm R}[k]\overline{\bm{\Xi}}[k]\overline{\mathbf{A}}_{\rm U}^H[k]$, where $\overline{\mathbf{B}}_{\rm R}[k]\triangleq[\mathbf{b}_{\rm R}(f_k,r_{R,1}),\cdots,\mathbf{b}_{\rm R}(f_k,r_{R,P})]$, $\overline{\bm{\Xi}}[k]\triangleq\sqrt{\frac{N_UN_R}{P}}{\rm diag}([\beta_1e^{-j2\pi\tau_1f_k},\cdots,\beta_Pe^{-j2\pi\tau_Pf_k}])$, and $\overline{\mathbf{A}}_{\rm U}[k]\triangleq[\mathbf{a}_{\rm U}(f_k,\widetilde{\vartheta}_{U,1},\widetilde{\varphi}_{U,1}),\cdots,\mathbf{a}_{\rm U}(f_k,\widetilde{\vartheta}_{U,P},\widetilde{\varphi}_{U,P})]$.
 
The 2D-OLS solution, with the known response vectors $\overline{\mathbf{B}}_{\rm R}[k]$ and $\overline{\mathbf{A}}_{\rm U}[k]$,
 is given by
  \begin{equation}
	\widehat{\overline{\bm{\Xi}}}[k]=(\overline{\bm{\Phi}}_{\rm R}^H[k]\overline{\bm{\Phi}}_{\rm R}[k])^{-1}\overline{\bm{\Phi}}_{\rm R}^H[k]{\mathbf{Y}}[k]\overline{\bm{\Phi}}_{\rm U}[k](\overline{\bm{\Phi}}_{\rm U}^H[k]\overline{\bm{\Phi}}_{\rm U}[k])^{-1},
\end{equation}
where $\overline{\bm{\Phi}}_{\rm R}[k]\triangleq \widetilde{\mathbf{V}}[k]\overline{\mathbf{B}}_{\rm R}[k]$ and $\overline{\bm{\Phi}}_{\rm U}[k]\triangleq \mathbf{F}^H[k]\overline{\mathbf{A}}_{\rm U}[k]$.
Hence, the reconstructed channel can be written as $\widehat{\overline{\mathbf{H}}}_{\rm U}[k] \triangleq \overline{\mathbf{B}}_{\rm R}[k]\widehat{\overline{\bm{\Xi}}}[k]\overline{\mathbf{A}}_{\rm U}^H[k]$.
The mean NMSE of 2D-OLS is given by
\begin{equation}\label{NMSEO}
	\text{NMSE}_\text{2D-OLS}\triangleq	\mathbb{E}\left\{\frac{1}{K}\sum_{k=1}^{K}\frac{\left\Vert  \mathbf{H}_{\rm U}[k]-\widehat{\overline{\mathbf{H}}}_{\rm U}[k] \right\Vert_F^2} { \left\Vert  \mathbf{H}_{\rm U}[k] \right\Vert_F^2}\right\}.
\end{equation}
The following proposition discusses its lower bound.
 \begin{Proposition}\label{proposition6}
With the 2D-OLS solution,
the mean NMSE is lower-bounded as 
	\begin{equation}\label{NMOLS}
	\text{NMSE}_\text{2D-OLS}\geq\frac{K\sigma_n^2P^5}{\kappa N_RN_U\sum_{k=1}^{K}\gamma_k},
	\end{equation}
where $\gamma_k\triangleq \frac{\bar{\lambda}_{\rm max}\left(\overline{\mathbf{B}}_{\rm R}^H[k]\overline{\mathbf{B}}_{\rm R}[k]\right)\bar{\lambda}_{\rm max}\left(\overline{\mathbf{A}}_{\rm U}^H[k]\overline{\mathbf{A}}_{\rm U}[k]\right)\left\Vert \overline{\bm{\Phi}}_{\rm U}[k]\right\Vert_F^2\left\Vert \overline{\bm{\Phi}}_{\rm R}[k]\right\Vert_F^2}{\bar{\lambda}_{\rm min}\left(\overline{\mathbf{B}}_{\rm R}^H[k]\overline{\mathbf{B}}_{\rm R}[k]\right)\bar{\lambda}_{\rm min}\left(\overline{\mathbf{A}}_{\rm U}^H[k]\overline{\mathbf{A}}_{\rm U}[k]\right)} $, and $\kappa\triangleq \sum_{p=1}^{P}\vert\beta_p\vert^2$ is the total path gain of the RIS-user channel.
\end{Proposition}
\begin{Proof}
 The mean MSE of the 2D-OLS estimator is expressed as
 \begin{equation}
 	\begin{aligned}
 			\mathbb{E}\left\{\left\Vert  \mathbf{H}_{\rm U}[k]-\widehat{\mathbf{H}}_{\rm U}[k] \right\Vert_F^2\right\}
 		=&\mathbb{E}\left\{\left\Vert  \overline{\mathbf{B}}_{\rm R}[k]\left({\overline{\bm{\Xi}}}[k]-\widehat{\overline{\bm{\Xi}}}[k]\right)\overline{\mathbf{A}}_{\rm U}^H[k] \right\Vert_F^2\right\} \\
 		\ge&  \bar{\lambda}_{\rm min}\left(\overline{\mathbf{B}}_{\rm R}^H[k]\overline{\mathbf{B}}_{\rm R}[k]\right)\bar{\lambda}_{\rm min}\left(\overline{\mathbf{A}}_{\rm U}^H[k]\overline{\mathbf{A}}_{\rm U}[k]\right)\mathbb{E}\left\{\left\Vert {\overline{\bm{\Xi}}}[k]-\widehat{\overline{\bm{\Xi}}}[k]\right\Vert_F^2 \right\},
 	\end{aligned} 
 \end{equation}
 where $\bar{\lambda}_{\rm min}(\mathbf{X})$ is the minimum eigenvalue of matrix $\mathbf{X}$. 
 Furthermore, we have
 \begin{equation}\label{EX}
 	\begin{aligned}
 \mathbb{E}\left\{	\left\Vert {\overline{\bm{\Xi}}}[k]-\widehat{\overline{\bm{\Xi}}}[k]\right\Vert_F^2 \right\} =&\mathbb{E}\left\{\left\Vert (\overline{\bm{\Phi}}_{\rm R}^H[k]\overline{\bm{\Phi}}_{\rm R}[k])^{-1}\overline{\bm{\Phi}}_{\rm R}^H[k]\widetilde{\mathbf{N}}[k]\overline{\bm{\Phi}}_{\rm U}[k](\overline{\bm{\Phi}}_{\rm U}^H[k]\overline{\bm{\Phi}}_{\rm U}[k])^{-1}
 	\right\Vert_F^2 \right\} \\
 	\overset{(b)}{=}&  \mathbb{E}\left\{ {\rm Tr} \left\{\mathbf{J}^H_{\rm R}[k]
 	\widetilde{\mathbf{N}}[k]\mathbf{J}_{\rm U}[k]\left(\mathbf{J}^H_{\rm R}[k]
 	\widetilde{\mathbf{N}}[k]\mathbf{J}_{\rm U}[k]\right)^H
 	\right\}\right\}  \\
 	=& {\rm Tr}\left\{ \mathbf{J}_{\rm R}^H[k] \mathbb{E}\left\{
 	\widetilde{\mathbf{N}}[k]\mathbf{J}_{\rm U}[k] \mathbf{J}^H_{\rm U}[k]
 	\widetilde{\mathbf{N}}^H[k]\right\}\mathbf{J}_{\rm R}[k] 
 	\right\} \\
 	\overset{(c)}{=}& \sigma_n^2{\rm Tr}\left\{  \mathbf{J}^H_{\rm U}[k] \mathbf{J}_{\rm U}[k]\right\}{\rm Tr}\left\{  \mathbf{J}^H_{\rm R}[k] \mathbf{J}_{\rm R}[k]\right\}  
 		\end{aligned}
 \end{equation}
where $(b)$ holds by defining
$\mathbf{J}_{\rm R}[k]\triangleq\overline{\bm{\Phi}}_{\rm R}(\overline{\bm{\Phi}}_{\rm R}^H[k]\overline{\bm{\Phi}}_{\rm R}[k])^{-1}$ and $\mathbf{J}_{\rm U}[k]\triangleq\overline{\bm{\Phi}}_{\rm U}[k](\overline{\bm{\Phi}}_{\rm U}^H[k]\overline{\bm{\Phi}}_{\rm U}[k])^{-1}$, and $(c)$  is established by the following derivation.
\begin{equation}\label{EN}
	\begin{aligned}
\mathbb{E}\left\{
\widetilde{\mathbf{N}}[k]\mathbf{J}_{\rm U}[k] \mathbf{J}^H_{\rm U}[k]
\widetilde{\mathbf{N}}^H[k]\right\}=&\mathbb{E}\left\{ \begin{bmatrix}
\widetilde{\mathbf{n}}_{1} \\ \widetilde{\mathbf{n}}_{2} \\ \vdots \\
\widetilde{\mathbf{n}}_{Q}
\end{bmatrix} \mathbf{J}_{\rm U}[k] \mathbf{J}^H_{\rm U}[k]\begin{bmatrix}
\widetilde{\mathbf{n}}_{1}^H & \widetilde{\mathbf{n}}_{2}^H & \cdots &
\widetilde{\mathbf{n}}_{Q}^H
\end{bmatrix}
\right\} \\
=&  \begin{bmatrix}
	\mathbb{E}\left\{\widetilde{\mathbf{n}}_{1}[k] \mathbf{J}_{\rm U}[k] \mathbf{J}^H_{\rm U}[k] 	\widetilde{\mathbf{n}}_{1}^H[k] \right\}&\cdots & 		\mathbb{E}\left\{\widetilde{\mathbf{n}}_{1}[k] \mathbf{J}_{\rm U}[k] \mathbf{J}^H_{\rm U}[k] 	\widetilde{\mathbf{n}}_{Q}^H[k]\right\} \\   \vdots & \ddots & \vdots  \\
	\mathbb{E}	\left\{\widetilde{\mathbf{n}}_{Q}[k] \mathbf{J}_{\rm U}[k] \mathbf{J}^H_{\rm U}[k] 	\widetilde{\mathbf{n}}_{1}^H[k]\right\} &\cdots & 		\mathbb{E}\left\{\widetilde{\mathbf{n}}_{Q}[k] \mathbf{J}_{\rm U}[k] \mathbf{J}^H_{\rm U}[k] 	\widetilde{\mathbf{n}}_{Q}^H[k] \right\}
\end{bmatrix} \\
\overset{(d)}{=} &\begin{bmatrix}
\sigma_n^2{\rm Tr}\left\{ \mathbf{J}_{\rm U}[k] \mathbf{J}^H_{\rm U}[k] \right\}&\cdots & 		0\\   \vdots & \ddots & \vdots  \\
0 &\cdots & 	\sigma_n^2	{\rm Tr}\left\{ \mathbf{J}_{\rm U}[k] \mathbf{J}^H_{\rm U}[k] \right\}
\end{bmatrix} ,
	\end{aligned}
\end{equation}
where the derivation of $(d)$ can be found in Appendix \ref{appendixC}.

Using the HM-GM-AM-QM inequalities, i.e., $\frac{\sum_{n=1}^{N}x_n}{N}\geq\frac{N}{\sum_{n=1}^{N}\frac{1}{x_n}}$, Eqn. (\ref{EX}) is simplified as
\begin{equation}\label{HM}
	\begin{aligned}
	\sigma_n^2{\rm Tr}\left\{  \mathbf{J}^H_{\rm U}[k] \mathbf{J}_{\rm U}[k]\right\}{\rm Tr}\left\{  \mathbf{J}^H_{\rm R}[k] \mathbf{J}_{\rm R}[k]\right\} =&	\sigma_n^2\sum_{p=1}^{P}\bar{\lambda}_{p}\left(\left(\overline{\bm{\Phi}}_{\rm U}^H[k]\overline{\bm{\Phi}}_{\rm U}[k]\right)^{-1} \right) \sum_{p=1}^{P}\bar{\lambda}_{p}\left(\left(\overline{\bm{\Phi}}_{\rm R}^H[k]\overline{\bm{\Phi}}_{\rm R}[k]\right)^{-1} \right)\\
	\geq &	\sigma_n^2 \frac{P^2}{\sum_{p=1}^{P}\frac{1}{\bar{\lambda}_{p}\left(\left(\overline{\bm{\Phi}}_{\rm U}^H[k]\overline{\bm{\Phi}}_{\rm U}[k]\right)^{-1} \right)}} \frac{P^2}{\sum_{p=1}^{P}\frac{1}{\bar{\lambda}_{p}\left(\left(\overline{\bm{\Phi}}_{\rm R}^H[k]\overline{\bm{\Phi}}_{\rm R}[k]\right)^{-1} \right)}} \\
	=&	\sigma_n^2 \frac{P^2}{\sum_{p=1}^{P}{\bar{\lambda}_{p}\left(\overline{\bm{\Phi}}_{\rm U}^H[k]\overline{\bm{\Phi}}_{\rm U}[k]\right) }} \frac{P^2}{\sum_{p=1}^{P}{\bar{\lambda}_{p}\left(\overline{\bm{\Phi}}_{\rm R}^H[k]\overline{\bm{\Phi}}_{\rm R}[k]\right) }} \\
	=&	\frac{\sigma_n^2P^4}{\left\Vert \overline{\bm{\Phi}}_{\rm U}[k]\right\Vert_F^2\left\Vert \overline{\bm{\Phi}}_{\rm R}[k]\right\Vert_F^2},
	\end{aligned}
\end{equation}
where  $\bar{\lambda}_p(\mathbf{X})$ denotes the $p$-th eigenvalue of matrix $\mathbf{X}$.

On the other hand, $\mathbf{H}_{\rm U}[k]$ is written by 
\begin{equation}\label{HUK}
	\begin{aligned}
	 \left\Vert  \mathbf{H}_{\rm U}[k] \right\Vert_F^2 &\leq  \bar{\lambda}_{\rm max}\left(\overline{\mathbf{B}}_{\rm R}^H[k]\overline{\mathbf{B}}_{\rm R}[k]\right)\bar{\lambda}_{\rm max}\left(\overline{\mathbf{A}}_{\rm U}^H[k]\overline{\mathbf{A}}_{\rm U}[k]\right)\mathbb{E}\left\{\left\Vert {\overline{\bm{\Xi}}}[k] \right\Vert_F^2 \right\} \\
	 & = \bar{\lambda}_{\rm max}\left(\overline{\mathbf{B}}_{\rm R}^H[k]\overline{\mathbf{B}}_{\rm R}[k]\right)\bar{\lambda}_{\rm max}\left(\overline{\mathbf{A}}_{\rm U}^H[k]\overline{\mathbf{A}}_{\rm U}[k]\right)\frac{N_RN_U\sum_{p=1}^{P}\vert\beta_p\vert^2}{P},
	\end{aligned}
\end{equation}
where $\bar{\lambda}_{\rm max}(\mathbf{X})$ is the maximum eigenvalue of matrix $\mathbf{X}$.

Thus, combining Eqns. (\ref{NMSEO}), (\ref{EX}), (\ref{HM}), and (\ref{HUK}), we have
	\begin{equation}\label{LB}
		\begin{aligned} 
	\text{NMSE}_\text{2D-OLS}&\geq  \frac{1}{K}\sum_{k=1}^{K}\frac{\bar{\lambda}_{\rm min}\left(\overline{\mathbf{B}}_{\rm R}^H[k]\overline{\mathbf{B}}_{\rm R}[k]\right)\bar{\lambda}_{\rm min}\left(\overline{\mathbf{A}}_{\rm U}^H[k]\overline{\mathbf{A}}_{\rm U}[k]\right)}{\bar{\lambda}_{\rm max}\left(\overline{\mathbf{B}}_{\rm R}^H[k]\overline{\mathbf{B}}_{\rm R}[k]\right)\bar{\lambda}_{\rm max}\left(\overline{\mathbf{A}}_{\rm U}^H[k]\overline{\mathbf{A}}_{\rm U}[k]\right)}\frac{ \sigma_n^2P^5}{N_RN_U\kappa\left\Vert \overline{\bm{\Phi}}_{\rm U}[k]\right\Vert_F^2\left\Vert \overline{\bm{\Phi}}_{\rm R}[k]\right\Vert_F^2} \\
	&=\frac{1}{K}\sum_{k=1}^{K}\frac{\sigma_n^2P^5}{N_RN_U\kappa\gamma_k}\\
	&\geq \frac{K\sigma_n^2P^5}{\kappa N_RN_U\sum_{k=1}^{K}\gamma_k}.
		\end{aligned}
\end{equation}
where $\gamma_k\triangleq \frac{\bar{\lambda}_{\rm max}\left(\overline{\mathbf{B}}_{\rm R}^H[k]\overline{\mathbf{B}}_{\rm R}[k]\right)\bar{\lambda}_{\rm max}\left(\overline{\mathbf{A}}_{\rm U}^H[k]\overline{\mathbf{A}}_{\rm U}[k]\right)\left\Vert \overline{\bm{\Phi}}_{\rm U}[k]\right\Vert_F^2\left\Vert \overline{\bm{\Phi}}_{\rm R}[k]\right\Vert_F^2}{\bar{\lambda}_{\rm min}\left(\overline{\mathbf{B}}_{\rm R}^H[k]\overline{\mathbf{B}}_{\rm R}[k]\right)\bar{\lambda}_{\rm min}\left(\overline{\mathbf{A}}_{\rm U}^H[k]\overline{\mathbf{A}}_{\rm U}[k]\right)} $, and $\kappa\triangleq \sum_{p=1}^{P}\vert\beta_p\vert^2$.

The proof of Prop. \ref{proposition6} is complete.
\end{Proof}
 \begin{remark}
 	The lower bound of 2D-OLS is comprised of two parts, one is the system hyperparameter $\frac{K\sigma_n^2P^5}{\kappa N_RN_U}$, and the other is $\sum_{k=1}^{K}\gamma_k$ related to the sensing matrix. Importantly,	
the lower bound can be further reduced by assuming the sensing matrix is semi-orthogonal such that $\gamma_k$ maximizes, indicating that the estimation performance can be improved by optimizing the sensing matrix.
This assumption needs to be supported by measurement matrix optimization\footnote{Given the designed dictionary, sensing matrix optimization is equivalent to measurement matrix optimization.}, which is left in the future work.
 \end{remark}

\section{Simulation Results}\label{simulation}
This section conducts several simulations to assess the efficacy of our proposed methods.  
The system setup considered for our simulations consists of $N_R^y=128$, $N_R^z=4$, $N_U^y=8$, $N_U^z=4$, $N_B^y=N_B^z=16$ and system frequency $f=28$ GHz with bandwidth $f_s=2$ GHz.  The noise power is set to $\sigma_n^2=-90$ dBm. 
For channel parameters, we employ the $28$ GHz mmWave channel setting in \cite{mmWave-channel}. Following Table. I in \cite{mmWave-channel}, we assume that that path gains follow $\mathcal{CN}(0,\aleph10^{-0.1{\rm PL}})$, where $\aleph=K_1^{1.8}10^{0.1K_2}$ with $K_1\sim \mathcal{U}(0,1)$ and $K_2\sim \mathcal{CN}(0,16)$. Besides, ${\rm PL}=a_1+10a_2{\rm log}_{10}(d^\prime)+\mathcal{N}(0,\sigma) [\rm{dB}]$ with $d^\prime$ denoting the distance in meters, where $a_1=72$, $a_2=2.92$ and $\sigma=8.7$ for NLoS paths, and $a_1=61.4$, $a_2=2$ and $\sigma=5.8$ for LoS paths. The distance between the BS and the RIS is set to $45$ meters. The users are uniformly distributed  around RIS 2 at a distance of $5$ to $20$ meters.  Moreover, we assume the number of paths $P=3$ (one is the LoS path) and the channel paths are uniformly distributed in elevation and azimuth coverage. The users' uplink power $\sigma_p^2$ is identical. \textcolor{black}{All the measurement matrices, i.e., the RIS phase shifts and the user's training matrix, are generated by making each element of them follow
	the complex Gaussian distribution $\mathcal{CN}(0,1)$ with modulus 1.}  
Lastly, the size of dictionaries is set as $G_R^y=N_R^y$, $G_R^z=8N_R^z$,  $G_U^y=8N_U^y$, and $G_U^z=8N_U^z$. All the stepsizes $\{\Delta_\theta,\Delta_\phi,\Delta_\vartheta,\Delta_\varphi,\Delta_r\}$ of the refined procedure are set as $0.005$.
The methods used for simulations are list as follows:
\begin{itemize}
	\item \textbf{2D-LS}: The wideband channel is estimated by Prop. \ref{2D-LSest}.
	\item \textbf{K-OMP}: The wideband channel is estimated by solving problem (P1) with OMP.
	\item \textbf{CC-MMPSR}: The wideband channel is estimated by our proposed MMPSR framework with CC-based atom matching.
	\item \textbf{IN-MMPSR}: The wideband channel is estimated by our proposed MMPSR framework with IN-based atom matching.
	\item \textbf{2D-OLS}: The wideband channel is estimated by the 2D-OLS estimator with the perfect channel support.  
	\item \textbf{LB}: The lower bound of the 2D-OLS estimator, which is derived in Section. \ref{LBD}.
\end{itemize}
 \vspace{-0.33cm}
\subsection{Evaluation of Different Methods}
\begin{figure}[htbp]
	\centering
	\subfigure[NMSE performance at $\sigma_p^2$ = 15 dBm]{
		\includegraphics[width=3.02in]{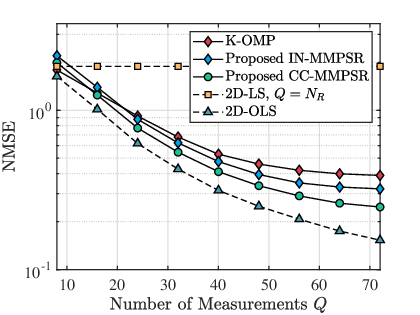}
	}
	\quad    
	\subfigure[NMSE performance at $\sigma_p^2$ = 30 dBm]{
		\includegraphics[width=3.02in]{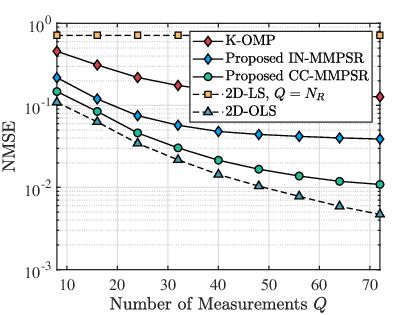}
	}
	\caption{NMSE versus the number of measurements $Q$ of different methods.}
	\label{CCIN}
\end{figure}
As analyzed in Section. \ref{NCE}, our framework shows that CC-based atom matching outperforms IN-based atom matching.  To support our findings, we conducted simulations with $N_X$ set to $32$, $Q$ ranging from $8$ to $72$, $K$ set to $32$, and the uplink power $\sigma_p^2$ set to either $15$ or $30$ dBm. Fig. \ref{CCIN} displays the NMSE performance plotted against the number of measurements $Q$. Our analysis reveals that the 2D-LS method performs poorly despite its high pilot overhead ($Q=N_R$ for 2D-LS). Instead, CC-MMPSR and IN-MMPSR are more effective, outperforming 2D-LS. Notably, CC-MMPSR performs better than IN-MMPSR in both low and high power scenarios, due to the fact that the 2-norm of each atom is typically unequal, which can interfere with the phase matching when using the IN with the target vector.
It should be noted that the performance improvement observed in the simulation results begins to plateau beyond a certain number of measurements. This is due to the fact that beyond a certain threshold, the additional measurements do not provide significant additional benefit to the accuracy of the channel estimation, while also adding complexity to the system.
 \vspace{-0.33cm}
\subsection{Impact of the Number of Subcarriers $K$}
\begin{figure*}
	\centering
	\subfigure[$K=1$]{
		\includegraphics[width=0.309\textwidth]{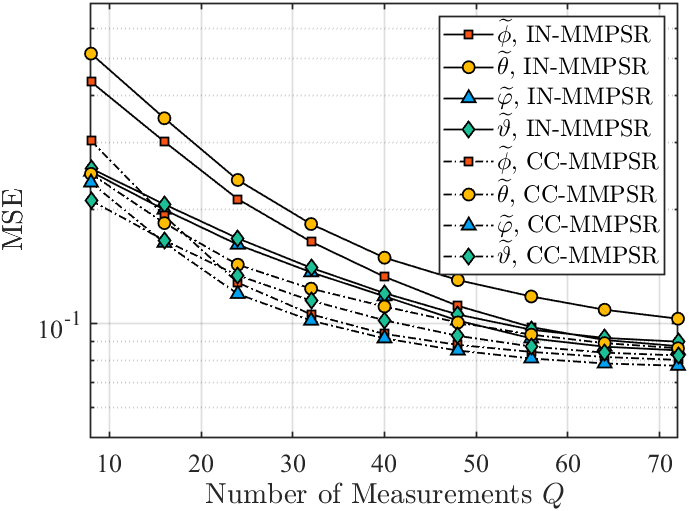}
	}
	\subfigure[$K=32$]{
		\includegraphics[width=0.309\textwidth]{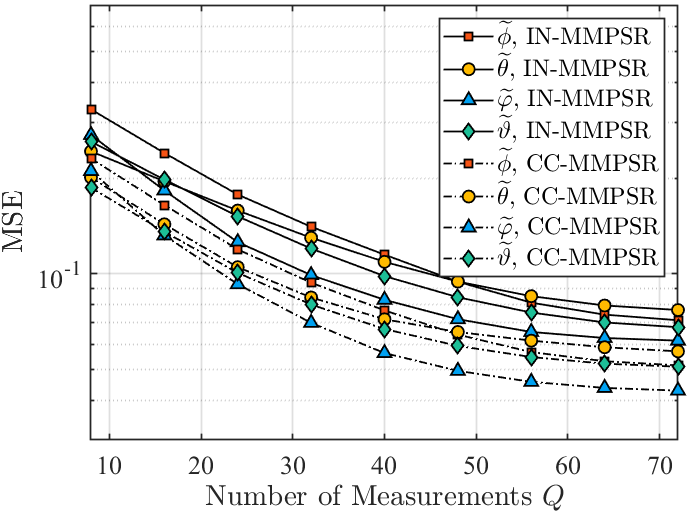}
	}
	\subfigure[$K=128$]{
		\includegraphics[width=0.309\textwidth]{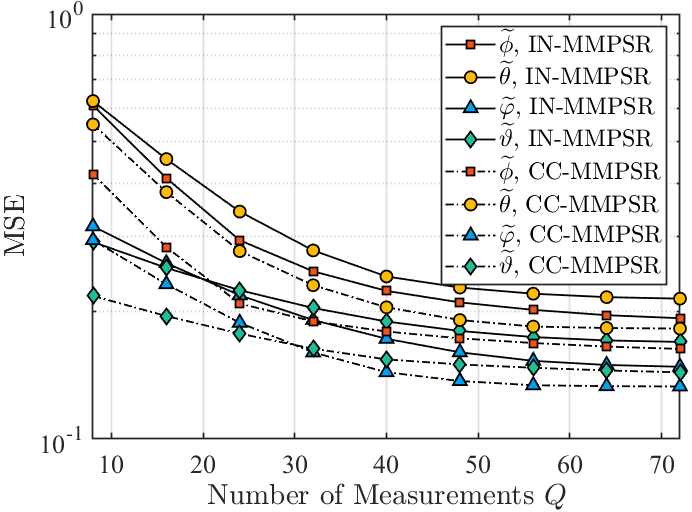}
	}
	\caption{MSE of the angles estimated by our proposed IN-MMPSR and CC-MMPSR.}
	\label{KANG}
\end{figure*}

\begin{figure}
	\centering
	\includegraphics[width = 0.49\textwidth]{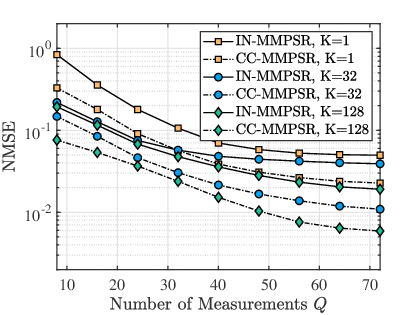}
	\caption{NMSE of the channel estimated by our proposed IN-MMPSR and CC-MMPSR.}
	\label{KCE}
\end{figure}

To evaluate the effectiveness of the proposed method that exploits the common sparsity support, Fig. \ref{KANG} and Fig. \ref{KCE} exhibit the MSE of the estimated angles\footnote{The estimated distance $\widehat{r}\gg 1$ is not exhibited here since it is not suitable for measuring with MSE.} and the NMSE of the estimated channels, respectively. The parameter setting includes that $\sigma_p^2=30$ dBm, $N_X=32$, $Q$ ranges from $8$ to $72$, and $K\in\{1,32,128\}$.   
The results indicate that CC-MMPSR outperforms IN-MMPSR in terms of MSE/NMSE for different numbers of subcarriers, as shown in Fig. \ref{KANG} and Fig. \ref{KCE}. Moreover, the joint parameter estimation provided by IN-/CC-MMPSR can further enhance the performance, as indicated by the decreasing MSE/NMSE trend as $K$ increases.
 
\subsection{Impact of the Near-Field Effect}
\begin{figure}
 	\centering
 	\includegraphics[width = 0.49\textwidth]{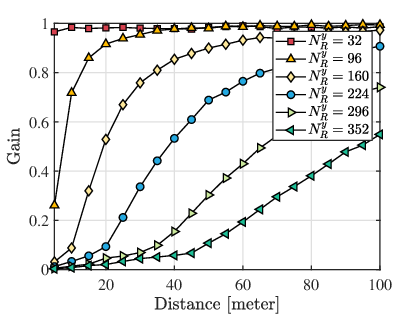}
 	\caption{The defined gain varies with the communication distance and the RIS size. }
 	\label{distance}
 \end{figure}

To demonstrate the significance of addressing the near-field effect, we generate a gain-distance curve for varying RIS sizes. The gain is defined by $\left\vert{\rm det}(\widetilde{\mathbf{B}}^H_{\rm R}\overline{\mathbf{B}}_{\rm R})\right\vert$, where $\widetilde{\mathbf{B}}_{\rm R}\in\mathbb{C}^{N_R\times P}$ is the planar channel support which has accurate angles but ignores the distance (the distance is assumed to be large enough).  Our simulation reveals two critical observations, as shown in Fig. \ref{distance}. Firstly, as the size of the RIS increases while keeping the distance constant, the gain decreases, primarily due to the near-field effect being exacerbated by the larger aperture. Secondly, when the size of the RIS is fixed and the distance between the RIS and the user increases, the gain also increases because the planar wavefront assumption holds in the far-field region. In essence, we demonstrate that the distance and array aperture play a significant role in reflecting the near-field effect. 
 \vspace{-0.33cm}
\subsection{Lower Bound Analysis} 
\begin{figure}[htbp]
	\centering
	\subfigure[NMSE versus $Q$]{
		\includegraphics[width=3.02in]{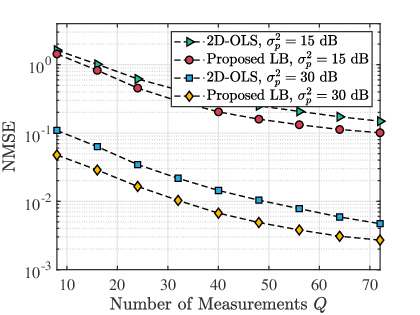}
	}
	\quad    
	\subfigure[NMSE versus $\sigma_p^2$]{
		\includegraphics[width=3.02in]{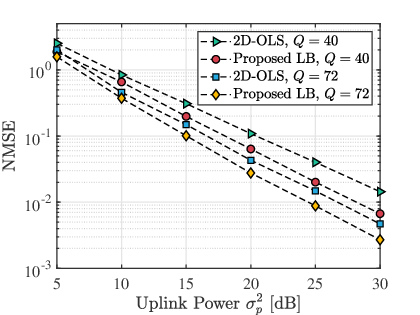}
	}
	\caption{The NMSE performance varies with the number of measurements and the uplink power.}
	\label{LBQP}
\end{figure}

To demonstrate the efficacy of the proposed lower bound, we compare 2D-OLS and LB under various simulation parameters. In Fig. \ref{LBQP}, we analyze the lower bound presented in Section \ref{LBD} using the following parameter settings: $K=32$, $\sigma_p^2$ in the range of $15$ to $30$ dB, $Q$ ranging from $8$ to $72$ for Fig. \ref{LBQP}(a), and $K=32$, $Q$ ranging from $40$ to $72$ at uplink power of $5$ to $30$ dB for Fig. \ref{LBQP}(b). The figure illustrates that the proposed LB achieves lower NMSE compared to 2D-OLS for various uplink power and number of measurements.
 \vspace{-0.18cm}
\subsection{Time Complexity Analysis}
First, the complexity of 1D-LS and 2D-LS is analyzed. According to Eqn. (\ref{1DLS}), the complexity is $\mathcal{O}(L^3N_U^3)$. 2D-LS of Eqn. (\ref{2DLS}), however, just incurs a complexity of $\mathcal{O}(L^3)$, where $L>N_U$ is considered.
The complexity of K-OMP with Eqn. (\ref{KCS}), dominated by atom matching, is $\mathcal{O}(PKQN_XG_UG_R)$.
For our proposed CC-MMPSR and IN-MMPSR, they have a comparable complexity due to the same linear complexity of IN and CC.  MMPSR consists of three steps: 1) SVD for the low-dimensional measurement matrix, 2) atom matching, and 3) parameter refinement. For step 1), the complexity is $\mathcal{O}\left(KQN_X{\rm min}(Q,N_X)\right)$, which could be reduced by some SVD methods used for sparse decomposition. The atom matching process, whether IN or CC, will incur a linear complexity of $\mathcal{O}(KG_RQ+KG_UN_X)$. For simplicity, the complexity of the refinement procedure can be designed to be smaller or comparable to that of the atom matching process. In XL-RIS systems, $G_R> N_X {\rm min}(Q,N_X)$ can be assumed such that the total complexity is determined by $\mathcal{O}(KG_RQ)$.
 \vspace{-0.5cm}
\section{Conclusions}\label{Con}
To summarize, this study has provided valuable insights into XL-RIS channel estimation in the wideband spherical domain. By analyzing the near-field beam squint effect and a 2D training signal structure, we have proposed a MMPSR recovery framework that is more efficient than the KCS framework for reconstructing the wideband channel. We have also introduced the CC as a substitute for the IN to measure the most relevant atom to the signal subspace and demonstrated through simulation that CC-MMPSR outperforms IN-MMPSR in terms of NMSE and time complexity.
Additionally, we have proposed the 2D-OLS estimator as a benchmark that cannot be surpassed by any 2D channel reconstruction method and derived its lower bound by considering all subcarriers, highlighting the significance of optimizing training matrices to reduce pilot overhead. However, this is challenging with near-field beam squint, and practical XL-RIS systems are greatly sensitive to the time complexity.
Therefore, future work will focus on proposing specific recovery frameworks for XL CS problems and exploring the use of deep learning to improve solutions.

 \vspace{-0.33cm}
\begin{appendices}
	 \vspace{-0.33cm}
	\section{ } \label{TM}
		According to Eqn. (\ref{Sk}), $g_y(\widetilde{\phi},r,k)$ is defined by
	\begin{equation}
		g_y(\widetilde{\phi},r,k)\triangleq \frac{1}{N_R^y} \left\vert \int_{-\frac{N_R^y}{2}}^{\frac{N_R^y}{2}} e^{j\frac{\pi d^2}{c}\left(f_k\frac{1-\widetilde{\phi}^2}{r}-f_c\frac{1-\widetilde{\phi}_{\bar{k}}^2}{r_{\bar{k}}} \right)(m_R^z)^2 }  {\rm d}m_R^z\right\vert.
	\end{equation}
	Denoting $\zeta_{\phi,k}\triangleq\frac{ d^2}{c}\left(f_k\frac{1-\widetilde{\phi}^2}{r}-f_c\frac{1-\widetilde{\phi}_{\bar{k}}^2}{r_{\bar{k}}} \right)$ for clarity, and then we define the equivalent function	
	\begin{equation}\label{wgy}
		\widetilde{g}_y(\zeta_{{\phi},k})\triangleq \frac{1}{N_R^y} \left\vert \int_{-\frac{N_R^y}{2}}^{\frac{N_R^y}{2}} e^{j{\pi}\zeta_{\phi,k}(m_R^z)^2 }  {\rm d}m_R^z\right\vert.
	\end{equation}	
	By substituting $j\pi\zeta_{\phi,k}(m_R^z)^2\triangleq -t^2$ into Eqn. (\ref{wgy}), we obtain
	\begin{equation}
		\begin{aligned}
			\widetilde{g}_y(\zeta_{{\phi},k})=&\left\vert  \frac{2(1+j)}{\sqrt{2\pi\zeta_{\phi,k}}N_R^y}  \int_{0}^{\frac{1-j}{2\sqrt{2}}\sqrt{\pi\zeta_{\phi,k}}N_{R}^y}e^{-t^2}{\rm d}t
			\right\vert
			\\
			=&\left\vert  \frac{1}{\sqrt{ \zeta_{\phi,k}}N_R^y}  {\rm erf}\left(
			\frac{1-j}{2\sqrt{2}}\sqrt{\pi\zeta_{\phi,k}}N_{R}^y
			\right) \right\vert,
		\end{aligned}
	\end{equation}	
	where the error function ${\rm erf}(x)\triangleq \frac{2}{\sqrt{\pi}}\int_{0}^{x}e^{-t^2}{\rm d}t$. 
	Similarly, $g_z(\widetilde{\theta},r,k) $ can be derived as 
	\begin{equation}
		\widetilde{g}_z(\zeta_{{\theta},k})=\left\vert  \frac{1}{\sqrt{ \zeta_{\theta,k}}N_R^z}  {\rm erf}\left(
		\frac{1-j}{2\sqrt{2}}\sqrt{\pi\zeta_{\theta,k}}N_{R}^z
		\right) \right\vert.
	\end{equation}	
	
	The proof is complete.

 \vspace{-0.43cm}
\section{  }\label{appendixB}
	Recalling Eqn. (\ref{Ykkk}), we know that $\mathcal{C}(\mathbf{Y}[k])\subset\mathcal{C}(\bm{\Phi}_{\rm R}[k]{\bm{\Xi}}[k])\subset\bm{\Phi}_{\rm R}[k]\mathcal{C}({\bm{\Xi}}[k])$, as well as $\mathcal{C}(\mathbf{Y}^H[k])\subset\mathcal{C}(\bm{\Phi}_{\rm U}[k]\bm{\Xi}^H[k])\subset\bm{\Phi}_{\rm U}[k]\mathcal{C}({\bm{\Xi}}^H[k])$.  $\bm{\xi}_{{\rm R},p}[k]$ and $\bm{\xi}_{{\rm U},p}[k]$ are the column vectors of the SVD of $\mathbf{Y}[k]$, such that $\bm{\xi}_{{\rm R},p}\in\mathcal{C}(\mathbf{Y}[k])\subset\bm{\Phi}_{\rm R}[k]\mathcal{C}({\bm{\Xi}}[k])$ and $\bm{\xi}_{{\rm U},p}[k]\in\mathcal{C}(\mathbf{Y}^H)\subset\bm{\Phi}_{\rm U}[k]\mathcal{C}({\bm{\Xi}}^H[k])$. Therefore, the solution of $ \underset{\bm{\xi}_{i,p}[k]}{{\rm min}}\{ \Vert\bm{\delta}_{i,k} \Vert_0+ \Vert\widetilde{\mathbf{t}}_{i,p}[k]-\bm{\Phi}_i[k]\bm{\xi}_{i,p}[k]\Vert_2^2\}$ is $P$-sparse due to $\bm{\Phi}_{\rm R}[k]\widehat{\bm{\xi}}_{{\rm R},p}[k]\approx\widetilde{\mathbf{t}}_{{\rm R},p}[k]\in\bm{\Phi}_{\rm R}[k]\mathcal{C}({\bm{\Xi}}[k])$ and $\bm{\Phi}_{\rm U}[k]\widehat{\bm{\xi}}_{{\rm U},p}[k]\approx\widetilde{\mathbf{t}}_{{\rm U},p}[k]\in\bm{\Phi}_{\rm U}[k]\mathcal{C}({\bm{\Xi}}^H[k])$. In particular, since our focus is on the most crucial component in each subspace, we are able to recover a single parameter in each subspace. In this context, each subspace corresponds to one channel support. Furthermore, the subspace signal can be jointly recovered via the common sparsity support of Eqn. (\ref{supp}).

The proof of Prop. \ref{svd} is completed.

 \vspace{-0.43cm}
\section{  }\label{appendixC} 
Here, the proof of $(d)$ in Eqn. (\ref{EN}) is given.

Consider a simplified form of $\widetilde{\mathbf{n}}_{q_1}\mathbf{X}\widetilde{\mathbf{n}}_{q_2}^H$, where $\forall q_1,q_2\in\{1,\cdots,Q\}$. 
 \begin{equation}
 	\widetilde{\mathbf{n}}_{q_1}\mathbf{X}\widetilde{\mathbf{n}}_{q_2}^H=\sum_{n_1=1}^{N_X}\sum_{n_2=1}^{N_X} \widetilde{n}_{q_1,n_1}\widetilde{n}_{q_2,n_2}^*[\mathbf{X}]_{n_1,n_2},
 \end{equation}
where $\widetilde{n}_{q_1,n_1}\triangleq[\widetilde{\mathbf{n}}_{q_1}]_{n_1}$ and $\widetilde{n}_{q_2,n_2}\triangleq[\widetilde{\mathbf{n}}_{q_2}]_{n_2}$.
 For $\forall q_1,q_2,n_1,n_2$, $\mathbb{E}\{\widetilde{n}_{q_1,n_1}\widetilde{n}_{q_2,n_2}^*\}=\sigma_n^2$ if and only if $q_1=q_2,n_1=n_2$; otherwise,  $\mathbb{E}\{\widetilde{n}_{q_1,n_1}\widetilde{n}_{q_2,n_2}^*\}=0$. This indicates that $\mathbb{E}\{	\widetilde{\mathbf{n}}_{q_1}\mathbf{X}\widetilde{\mathbf{n}}_{q_2}^H\}$ equals to $\sigma_n^2{\rm Tr}\{\mathbf{X}\}$ at $q_1=q_2$. Therefore, the equation $(d)$ can be proved.

\end{appendices}

\bibliographystyle{IEEEtran}
\bibliography{reference.bib}

\begin{thebibliography}{10}
\providecommand{\url}[1]{#1}
\csname url@samestyle\endcsname
\providecommand{\newblock}{\relax}
\providecommand{\bibinfo}[2]{#2}
\providecommand{\BIBentrySTDinterwordspacing}{\spaceskip=0pt\relax}
\providecommand{\BIBentryALTinterwordstretchfactor}{4}
\providecommand{\BIBentryALTinterwordspacing}{\spaceskip=\fontdimen2\font plus
\BIBentryALTinterwordstretchfactor\fontdimen3\font minus
  \fontdimen4\font\relax}
\providecommand{\BIBforeignlanguage}[2]{{%
\expandafter\ifx\csname l@#1\endcsname\relax
\typeout{** WARNING: IEEEtran.bst: No hyphenation pattern has been}%
\typeout{** loaded for the language `#1'. Using the pattern for}%
\typeout{** the default language instead.}%
\else
\language=\csname l@#1\endcsname
\fi
#2}}
\providecommand{\BIBdecl}{\relax}
\BIBdecl

\bibitem{XL1}
M.~Cui, Z.~Wu, Y.~Lu, X.~Wei, and L.~Dai, ``Near-field {MIMO} communications
  for {6G}: Fundamentals, challenges, potentials, and future directions,''
  \emph{IEEE Commun. Mag.}, vol.~61, no.~1, pp. 40--46, 2023.

\bibitem{HMIMO}
C.~Huang, S.~Hu, G.~C. Alexandropoulos, A.~Zappone, C.~Yuen, R.~Zhang, M.~D.
  Renzo, and M.~Debbah, ``Holographic {MIMO} surfaces for 6g wireless networks:
  Opportunities, challenges, and trends,'' \emph{IEEE Wireless Commun.},
  vol.~27, no.~5, pp. 118--125, 2020.

\bibitem{OAM}
R.~Chen, M.~Chen, X.~Xiao, W.~Zhang, and J.~Li, ``Multi-user orbital angular
  momentum based {Terahertz} communications,'' \emph{IEEE Trans. Wireless
  Commun.}, pp. 1--1, 2023.

\bibitem{XL2}
H.~Zhang, N.~Shlezinger, F.~Guidi, D.~Dardari, M.~F. Imani, and Y.~C. Eldar,
  ``Beam focusing for near-field multiuser {MIMO} communications,'' \emph{IEEE
  Trans. Wireless Commun.}, vol.~21, no.~9, pp. 7476--7490, 2022.

\bibitem{XL3}
J.~Tian, Y.~Han, S.~Jin, and M.~Matthaiou., ``Low-overhead localization and vr
  identification for subarray-based {ELAA} systems,'' \emph{IEEE Wireless
  Commun. Lett.}, pp. 1--1, 2023.

\bibitem{XL4}
Z.~Lu, Y.~Han, S.~Jin, M.~Matthaiou, and T.~Q.~S. Quek, ``Near-field channel
  reconstruction and user localization for {ELAA} systems,'' in \emph{2022
  International Symposium on Wireless Communication Systems (ISWCS)}, 2022, pp.
  1--6.

\bibitem{near-field1}
K.~T. Selvan and R.~Janaswamy, ``Fraunhofer and fresnel distances : Unified
  derivation for aperture antennas.'' \emph{IEEE Antennas Propag. Mag.},
  vol.~59, no.~4, pp. 12--15, 2017.

\bibitem{SMART}
M.~Di~Renzo, A.~Zappone, M.~Debbah, M.-S. Alouini, C.~Yuen, J.~de~Rosny, and
  S.~Tretyakov, ``Smart radio environments empowered by reconfigurable
  intelligent surfaces: How it works, state of research, and the road ahead,''
  \emph{IEEE J. Sel. Areas Commun.}, vol.~38, no.~11, pp. 2450--2525, 2020.

\bibitem{RIS1}
W.~Tang, M.~Z. Chen, X.~Chen, J.~Y. Dai, Y.~Han, M.~Di~Renzo, Y.~Zeng, S.~Jin,
  Q.~Cheng, and T.~J. Cui, ``Wireless communications with reconfigurable
  intelligent surface: Path loss modeling and experimental measurement,''
  \emph{IEEE Trans. Wireless Commun.}, vol.~20, no.~1, pp. 421--439, 2021.

\bibitem{RIS2}
R.~Li, B.~Guo, M.~Tao, Y.-F. Liu, and W.~Yu, ``Joint design of hybrid
  beamforming and reflection coefficients in {RIS}-aided mmwave {MIMO}
  systems,'' \emph{IEEE Trans. Commun.}, vol.~70, no.~4, pp. 2404--2416, 2022.

\bibitem{FRIS1}
L.~Wei, C.~Huang, Q.~Guo, Z.~Yang, Z.~Zhang, G.~C. Alexandropoulos, M.~Debbah,
  and C.~Yuen, ``Joint channel estimation and signal recovery for
  {RIS}-empowered multiuser communications,'' \emph{IEEE Trans. Commun.},
  vol.~70, no.~7, pp. 4640--4655, 2022.

\bibitem{FRIS2}
S.~Yang, W.~Lyu, D.~Wang, and Z.~Zhang, ``Separate channel estimation with
  hybrid {RIS}-aided multi-user communications,'' \emph{IEEE Trans. Veh.
  Technol.}, pp. 1--6, 2022.

\bibitem{FRIS3}
L.~Wei, C.~Huang, G.~C. Alexandropoulos, C.~Yuen, Z.~Zhang, and M.~Debbah,
  ``Channel estimation for {RIS}-empowered multi-user {MISO} wireless
  communications,'' \emph{IEEE Trans. Commun.}, vol.~69, no.~6, pp. 4144--4157,
  2021.

\bibitem{FRIS4}
C.~Huang, Z.~Yang, G.~C. Alexandropoulos, K.~Xiong, L.~Wei, C.~Yuen, Z.~Zhang,
  and M.~Debbah, ``Multi-hop {RIS}-empowered terahertz communications: A
  {DRL}-based hybrid beamforming design,'' \emph{IEEE J. Sel. Areas Commun.},
  vol.~39, no.~6, pp. 1663--1677, 2021.

\bibitem{FRIS6}
A.~Fascista, M.~F. Keskin, A.~Coluccia, H.~Wymeersch, and G.~Seco-Granados,
  ``Ris-aided joint localization and synchronization with a single-antenna
  receiver: Beamforming design and low-complexity estimation,'' \emph{IEEE J.
  Sel. Top. Signal Process.}, vol.~16, no.~5, pp. 1141--1156, 2022.

\bibitem{RIS3}
C.~Huang, A.~Zappone, G.~C. Alexandropoulos, M.~Debbah, and C.~Yuen,
  ``Reconfigurable intelligent surfaces for energy efficiency in wireless
  communication,'' \emph{IEEE Trans. Wireless Commun.}, vol.~18, no.~8, pp.
  4157--4170, 2019.

\bibitem{NB1}
B.~Ning, Z.~Chen, W.~Chen, Y.~Du, and J.~Fang, ``Terahertz multi-user massive
  {MIMO} with intelligent reflecting surface: Beam training and hybrid
  beamforming,'' \emph{IEEE Trans. Veh. Technol.}, vol.~70, no.~2, pp.
  1376--1393, 2021.

\bibitem{NB2}
B.~Ning, Z.~Chen, W.~Chen, and J.~Fang, ``Beamforming optimization for
  intelligent reflecting surface assisted {MIMO}: A sum-path-gain maximization
  approach,'' \emph{IEEE Wireless Commun. Lett.}, vol.~9, no.~7, pp.
  1105--1109, 2020.

\bibitem{XLRIS1}
E.~Björnson, Ã.~T. Demir, and L.~Sanguinetti, ``A primer on near-field
  beamforming for arrays and reconfigurable intelligent surfaces,'' in
  \emph{2021 55th Asilomar Conference on Signals, Systems, and Computers},
  2021, pp. 105--112.

\bibitem{XLRIS2}
N.~Decarli and D.~Dardari, ``Communication modes with large intelligent
  surfaces in the near field,'' \emph{IEEE Access}, vol.~9, pp.
  165\,648--165\,666, 2021.

\bibitem{XL-BT1}
\BIBentryALTinterwordspacing
X.~Wei, L.~Dai, Y.~Zhao, G.~Yu, and X.~Duan, ``Codebook design and beam
  training for extremely large-scale ris: Far-field or near-field?'' 2021.
  [Online]. Available: \url{https://arxiv.org/abs/2109.10143}
\BIBentrySTDinterwordspacing

\bibitem{XL-BT2}
G.~C. Alexandropoulos, V.~Jamali, R.~Schober, and H.~V. Poor, ``Near-field
  hierarchical beam management for ris-enabled millimeter wave multi-antenna
  systems,'' in \emph{2022 IEEE 12th Sensor Array and Multichannel Signal
  Processing Workshop (SAM)}, 2022, pp. 460--464.

\bibitem{XL-BT3}
Y.~Zhang, X.~Wu, and C.~You, ``Fast near-field beam training for extremely
  large-scale array,'' \emph{IEEE Wireless Commun. Lett.}, vol.~11, no.~12, pp.
  2625--2629, 2022.

\bibitem{XL-loc1}
O.~Rinchi, A.~Elzanaty, and M.-S. Alouini, ``Compressive near-field
  localization for multipath {RIS}-aided environments,'' \emph{IEEE Commun.
  Lett.}, vol.~26, no.~6, pp. 1268--1272, 2022.

\bibitem{XL-loc2}
D.~Dardari, N.~Decarli, A.~Guerra, and F.~Guidi, ``Los/nlos near-field
  localization with a large reconfigurable intelligent surface,'' \emph{IEEE
  Trans. Wireless Commun.}, vol.~21, no.~6, pp. 4282--4294, 2022.

\bibitem{XL-loc3}
Y.~Han, S.~Jin, C.-K. Wen, and T.~Q. Quek, ``Localization and channel
  reconstruction for extra large {RIS}-assisted massive {MIMO} systems,''
  \emph{IEEE J. Sel. Top. Signal Process.}, pp. 1--1, 2022.

\bibitem{XL-loc4}
\BIBentryALTinterwordspacing
Y.~Pan, C.~Pan, S.~Jin, and J.~Wang, ``Ris-aided near-field localization and
  channel estimation for the sub-terahertz system,'' 2022. [Online]. Available:
  \url{https://arxiv.org/abs/2208.11343}
\BIBentrySTDinterwordspacing

\bibitem{XL-loc5}
R.~Ghazalian, K.~Keykhosravi, H.~Chen, H.~Wymeersch, and R.~Jäntti,
  ``Bi-static sensing for near-field {RIS} localization,'' in \emph{GLOBECOM
  2022 - 2022 IEEE Global Communications Conference}, 2022, pp. 6457--6462.

\bibitem{XLRCE}
S.~Yang, W.~Lyu, Z.~Hu, Z.~Zhang, and C.~Yuen, ``Channel estimation for
  near-field {XL-RIS}-aided mmwave hybrid beamforming architectures,''
  \emph{IEEE Trans. Veh. Technol.}, pp. 1--6, 2023.

\bibitem{near-CE2}
M.~Cui and L.~Dai, ``Channel estimation for extremely large-scale {MIMO}:
  Far-field or near-field?'' \emph{IEEE Trans. Commun.}, vol.~70, no.~4, pp.
  2663--2677, 2022.

\bibitem{BSE5}
A.~M. Elbir, K.~V. Mishra, and S.~Chatzinotas, ``{NBA-OMP}: Near-field
  beam-split-aware orthogonal matching pursuit for wideband {THz} channel
  estimation,'' 2023.

\bibitem{BSE1}
X.~Gao, L.~Dai, S.~Zhou, A.~M. Sayeed, and L.~Hanzo, ``Wideband beamspace
  channel estimation for millimeter-wave {MIMO} systems relying on lens antenna
  arrays,'' \emph{IEEE Trans. Signal Process.}, vol.~67, no.~18, pp.
  4809--4824, 2019.

\bibitem{BSE2}
Y.~Chen, Y.~Xiong, D.~Chen, T.~Jiang, S.~X. Ng, and L.~Hanzo, ``Hybrid
  precoding for wideband millimeter wave {MIMO} systems in the face of beam
  squint,'' \emph{IEEE Trans. Wireless Commun.}, vol.~20, no.~3, pp.
  1847--1860, 2021.

\bibitem{BSE4}
S.~Ma, W.~Shen, J.~An, and L.~Hanzo, ``Wideband channel estimation for
  {IRS}-aided systems in the face of beam squint,'' \emph{IEEE Trans. Wireless
  Commun.}, vol.~20, no.~10, pp. 6240--6253, 2021.

\bibitem{RIS-NF-BSE1}
A.~Lee, H.~Ju, S.~Kim, and B.~Shim, ``Intelligent near-field channel estimation
  for {Terahertz} ultra-massive {MIMO} systems,'' in \emph{GLOBECOM 2022 - 2022
  IEEE Global Communications Conference}, 2022, pp. 5390--5395.

\bibitem{RIS-NF-BSE2}
\BIBentryALTinterwordspacing
H.~Luo and F.~Gao, ``Beam squint assisted user localization in near-field
  communications systems,'' 2022. [Online]. Available:
  \url{https://arxiv.org/abs/2205.11392}
\BIBentrySTDinterwordspacing

\bibitem{RIS-NF-BSE3}
\BIBentryALTinterwordspacing
J.~An, C.~Xu, D.~W.~K. Ng, C.~Yuen, L.~Gan, and L.~Hanzo, ``Reconfigurable
  intelligent surface-enhanced {OFDM} communications via delay adjustable
  metasurface,'' 2021. [Online]. Available:
  \url{https://arxiv.org/abs/2110.09291}
\BIBentrySTDinterwordspacing

\bibitem{XLRIS3}
\BIBentryALTinterwordspacing
W.~Hao, X.~You, F.~Zhou, Z.~Chu, G.~Sun, and P.~Xiao, ``The far-/near-field
  beam squint and solutions for thz intelligent reflecting surface
  communications,'' 2022. [Online]. Available:
  \url{https://arxiv.org/abs/2208.12385}
\BIBentrySTDinterwordspacing

\bibitem{Spherdic}
\BIBentryALTinterwordspacing
Z.~Wu and L.~Dai, ``Multiple access for near-field communications: {SDMA} or
  {LDMA}?'' 2022. [Online]. Available: \url{https://arxiv.org/abs/2208.06349}
\BIBentrySTDinterwordspacing

\bibitem{2DLSE}
\BIBentryALTinterwordspacing
\emph{Linear Systems and Characterization of Generalized Inverses}.\hskip 1em
  plus 0.5em minus 0.4em\relax New York, NY: Springer New York, 2003, pp.
  52--103. [Online]. Available: \url{https://doi.org/10.1007/0-387-21634-0_4}
\BIBentrySTDinterwordspacing

\bibitem{MMV}
J.~Chen and X.~Huo, ``Theoretical results on sparse representations of
  multiple-measurement vectors,'' \emph{IEEE Trans. Signal Process.}, vol.~54,
  no.~12, pp. 4634--4643, 2006.

\bibitem{SVD}
S.~Friedland, Q.~Li, and D.~Schonfeld, ``Compressive sensing of sparse
  tensors,'' \emph{IEEE Trans. Image Process.}, vol.~23, no.~10, pp.
  4438--4447, 2014.

\bibitem{OLS1}
Z.~Gao, L.~Dai, W.~Dai, B.~Shim, and Z.~Wang, ``Structured compressive
  sensing-based spatio-temporal joint channel estimation for {FDD} massive
  {MIMO},'' \emph{IEEE Trans. on Commun.}, vol.~64, no.~2, pp. 601--617, 2016.

\bibitem{OLS2}
J.~Lee, G.-T. Gil, and Y.~H. Lee, ``Channel estimation via orthogonal matching
  pursuit for hybrid {MIMO} systems in millimeter wave communications,''
  \emph{IEEE Trans. Commun.}, vol.~64, no.~6, pp. 2370--2386, 2016.

\bibitem{mmWave-channel}
G.~R. MacCartney, M.~K. Samimi, and T.~S. Rappaport, ``Omnidirectional path
  loss models in new york city at 28 {GHz} and 73 {GHz},'' in \emph{2014 IEEE
  25th Annual International Symposium on Personal, Indoor, and Mobile Radio
  Communication (PIMRC)}, 2014, pp. 227--231.

\end{thebibliography}

\vspace{12pt}

\end{document}